\documentclass[12pt]{article}
\usepackage{graphicx}
\usepackage{dcolumn}
\usepackage{bm}
\usepackage{amsmath, amssymb, amsfonts, amsthm}
\usepackage{cite, wrapfig}
\usepackage{epsfig}
\usepackage[usenames,dvipsnames,table]{xcolor}
\usepackage{textcomp}
\usepackage{tikz-cd} 
\usepackage{ dsfont }
\usepackage[utf8]{inputenc}
\usepackage{amsmath}
\usepackage{tikz}
\usepackage{youngtab}
\usepackage{hyperref}

\usepackage{colortbl}
\definecolor{summersky}{cmyk}{0.71,0.33,0,0.5}
\definecolor{flamingo}{cmyk}{0,0.51,0.71,0.5}
\definecolor{rp}{cmyk}{0.2, 1, 0.6, 0}
\definecolor{pacificblue}{cmyk}{0.95,0.3,0, 0.5}
\definecolor{gray60}{cmyk}{0.4,0.4,0,0.8}

\hypersetup{
    pdftoolbar=true,        
    pdfmenubar=true,        
    pdffitwindow=true,     
    pdfstartview={FitH},    
    pdfnewwindow=true,      
    colorlinks=true,       
    linkcolor=pacificblue,          
    citecolor=flamingo,        
    filecolor=magenta,      
    urlcolor=pacificblue           
}

\def\be{\begin{equation}}
\def\ee{\end{equation}}
\def\ba{\begin{eqnarray}}
\def\ea{\end{eqnarray}}

\def\bdm{\begin{displaymath}}
\def\edm{\end{displaymath}}

\def\bq{\begin{quote}}
	\def\eq{\end{quote}}

 at 10truept

\newcommand{\bea}{\begin{eqnarray}}
\newcommand{\eea}{\end{eqnarray}}

\newcommand{\bi}{\begin{itemize}}
	\newcommand{\ei}{\end{itemize}}

\newcommand{\beq}{\begin{equation}}
\newcommand{\eeq}{\end{equation}}
\newcommand{\beqa}{\begin{eqnarray}}
\newcommand{\eeqa}{\end{eqnarray}}

\def\ltap{\ \raise.3ex\hbox{$<$\kern-.75em\lower1ex\hbox{$\sim$}}\ }
\def\gtap{\ \raise.3ex\hbox{$>$\kern-.75em\lower1ex\hbox{$\sim$}}\ }
\def\gl{\ \raise.5ex\hbox{$>$}\kern-.8em\lower.5ex\hbox{$<$}\ }
\def\roughly#1{\raise.3ex\hbox{$#1$\kern-.75em\lower1ex\hbox{$\sim$}}}

\begin{document}

		\thispagestyle{empty}
		\begin{flushright}
			May 2019\\
		\end{flushright}
		\vspace*{.2cm}
		\begin{center}			
			{\Large \bf An Algebraic Classification of Exceptional EFTs \\ \vspace{0.2cm} Part II: Supersymmetry}
			
			\vspace*{.7cm} {\large Diederik Roest\footnote{\tt d.roest@rug.nl
				}, David Stefanyszyn\footnote{\tt
				d.stefanyszyn@rug.nl} and Pelle Werkman\footnote{\tt
				p.j.werkman@rug.nl}}

		\vspace{.3cm} {\em Van Swinderen Institute for Particle Physics and Gravity, University of Groningen, Nijenborgh 4, 9747 AG Groningen, The Netherlands}\\
\end{center}

		\vspace{1cm} 
We present a novel approach to classify supersymmetric effective field theories (EFTs) whose scattering amplitudes exhibit enhanced soft limits. These enhancements arise due to non-linearly realised symmetries on the Goldstone modes of such EFTs and we classify the algebras that these symmetries can form. Our main focus is on so-called \textit{exceptional} algebras which lead to field-dependent transformation rules and EFTs with the maximum possible soft enhancement at a given derivative power counting. We adapt existing techniques for Poincar\'{e} invariant theories to the supersymmetric case, and introduce \textit{superspace inverse Higgs constraints} as a method of reducing the number of Goldstone modes while maintaining all symmetries. 

Restricting to the case of a single Goldstone supermultiplet in four dimensions, we classify the exceptional algebras and  EFTs for a chiral, Maxwell or real linear supermultiplet. Moreover, we show how our algebraic approach allows one to read off the soft weights of the different component fields from \textit{superspace inverse Higgs trees}, which are the algebraic cousin of the on-shell soft data one provides to soft bootstrap EFTs using on-shell recursion. Our Lie-superalgebraic approach extends the results of on-shell methods and provides a complementary perspective on non-linear realisations.

	\vfill \setcounter{page}{0} \setcounter{footnote}{0}
	\newpage
	
	\tableofcontents
	

	\section{Introduction} \label{intro}
	 
Non-linear realisations of spontaneously broken symmetries are a central aspect of many areas of physics. We now have a very good understanding about the connection between non-linearly realised symmetries and the special infra-red (IR) behaviour of scattering amplitudes \cite{ScatteringAmplitudes}. The usual {lore} is that the symmetries are primary from which one can derive the corresponding soft theorems. However, the opposite approach has also proven fruitful: based on minimal assumptions regarding the linearly realised symmetries and soft theorems, one can construct amplitudes with special soft behaviour and derive the corresponding theories and symmetries. This soft bootstrap program has been applied to scalar effective field theories (EFTs) \cite{SoftLimits1,SoftLimits2,SoftLimits3,Recursion,LowYin}, vector EFTs \cite{SoftLimitVector} and supersymmetric EFTs \cite{SoftBootstrapSUSY, SuperGals} relying on new ideas \cite{Recursion} based on on-shell recursion techniques \cite{Recursion1,Recursion2, SimplestQFT}. In theories with constant shift symmetries one encounters Adler's zero \cite{Adler1,Adler2} while in theories with explicit coordinate dependent symmetries one encounters enhanced soft limits where soft amplitudes depend non-linearly on the soft momentum at leading order. This offers a very neat classification of EFTs which does not require any reference to Lagrangians or field bases. 

More specifically, if a theory is invariant under a symmetry transformation with a field-independent part with $\sigma-1$ powers of the space-time coordinates, then in the single soft limit where a single external momentum $p$ is taken soft, the amplitudes scale as $p^{\sigma}$ to leading order with $\sigma$ referred to as the soft weight\footnote{Note that this simple connection between symmetries and enhanced soft limits does not apply to gauge theories, where gauge symmetries can be thought of as an infinite number of coordinate dependent symmetries, but is certainly applicable to scalar and spin-$1/2$ fermions. We will comment on gauge theories in section \ref{Maxwell}.}. So theories with symmetries involving many powers of the coordinates decouple very quickly in the IR. This makes sense since the invariant operators would involve many derivatives which are suppressed at long wavelength. Note that we are assuming that the field-independent part of the symmetry transformation is compatible with a canonical propagator. This is important when understanding the soft behaviour of a dilaton, for example, where once we canonically normalise all terms in all transformation rules are field-dependent\footnote{The dilaton EFT non-linearly realises the conformal algebra so the symmetry transformations we refer to here are dilatations and special conformal transformations.}, see e.g. \cite{Mapping}. It does therefore not fit into the above classification but it is known that the dilaton has $\sigma = 0$ soft behaviour \cite{dilaton1,dilaton2,dilaton3}.

However, the soft amplitude bootstrap is not the only way of classifying these special EFTs without reference to Lagrangians. Any symmetries which are non-linearly realised on the fields must form a consistent Lie-algebra with the assumed linearly realised symmetries. One can therefore ask which Lie-algebras are consistent within the framework of the coset construction for non-linear realisations \cite{Internal1,Internal2,Spacetime1} augmented with the crucial inverse Higgs phenomenon\footnote{We note that in contrast to the case for internal symmetries, there is no proof of coset universality when space-time symmetries are spontaneously broken. In this work we will primarily be concerned with space-time symmetry breaking and will therefore assume that universality does hold.} \cite{Spacetime2}. For scalar EFTs Lie-algebraic approaches have been presented in \cite{LieAlgebraicScalar1,LieAlgebraicScalar2} while in \cite{LieAlgebraicVector} these methods were used to prove that a gauge vector cannot be a Goldstone mode of a spontaneously broken space-time symmetry without introducing new degrees of freedom. This implies that the Born-Infeld (BI) vector is not special from the perspective of non-linear symmetries and enhanced soft limits (the same result was found in \cite{SoftBootstrapSUSY} where it was shown that the BI vector has a vanishing soft weight).

Recently, we presented an algorithm  for an exhaustive classification of the possible algebras which can be non-linearly realised on a set of Goldstone modes with linearly realised Poincar\'{e} symmetries\footnote{See \cite{AdS} for a discussion on non-linearly realised symmetries in AdS/dS space-time rather than Minkowski space-time and \cite{SymmetricSuperfluids} for cases where Lorentz boosts are non-linearly realised.} and canonical propagators in \cite{RSW}. We illustrated this with EFTs of multiple scalars and multiple spin-$1/2$ fermions. A key aspect of this algorithm are inverse Higgs trees which incorporate the necessary requirements for the existence of inverse Higgs constraints in a systematic manner. These constraints arise when space-time symmetries are spontaneously broken and puts-into-practice the statement that Goldstone's theorem \cite{Goldstone} does not apply beyond the breaking of internal symmetries \cite{LowManohar}. Indeed, we can realise space-time symmetries on fewer Goldstones than broken generators, which underlies the existence of enhanced soft limits in special EFTs. The inverse Higgs tree can be seen as the algebraic cousin to the on-shell soft data one provides in the soft bootstrap program. Indeed, the tree encodes information about the massless states, linearly realised symmetries and soft weights.  Our algorithm allows one to   establish in a simple manner which generators can be included in a non-linearly realised algebra, given a set of Goldstone modes: only generators which live in a Taylor expansion of the Goldstone modes are consistent while the existence of canonical propagators restricts these generators further.

At the Lie-algebraic level there are two distinct types of algebras which are of interest. The first possibility has vanishing commutators between all non-linear generators (which correspond to spontaneously broken symmetries as opposed to linear generators which generate linearly realised symmetries) which leads to field-independent extended shift symmetries for the Goldstone modes \cite{ExtendedShift}. These are simply shift symmetries which are monomial in the space-time coordinates, with higher powers leading to quicker decoupling in the IR. In the resulting EFTs, the operators of most interest are the Wess-Zumino ones since these have fewer derivatives per field than the strictly invariant operators, of which the scalar Galileon interactions \cite{Galileon, WZ} are an important example.

The other possibility is to have at least one non-vanishing commutator between a pair of non-linear generators. This leads to field-dependent transformation rules for the Goldstones and exceptional EFTs. These are particularly interesting since the symmetry relates operators of different mass dimensions, most notably relating the propagator to leading order interactions. In terms of Feynman diagrams, the exceptional EFTs exhibit cancellations between pole and contact diagrams. 

A very well known example of an exceptional EFT is the scalar sector of the Dirac-Born-Infeld (DBI) action \cite{BI,Dirac} which describes the fluctuations of a probe brane in an extra dimension. A second possibility is the Special Galileon \cite{Galileon,SpecialGalileon} which has been studied from various directions \cite{DoubleCopy,GeometryGal}. In our recent paper \cite{RSW} we have demonstrated from an algebraic perspective that these are the only two exceptional algebras and EFTs for a single scalar field. Moreover, in the context of fermionic Goldstones, we proved that the only exceptional EFT is that corresponding to Volkov-Akulov (VA) \cite{VA} and its multi-field extensions which non-linearly realise supersymmetry (SUSY) algebras\footnote{Since the VA symmetry starts out with a constant shift, which is augmented with field-dependent pieces, it has a $\sigma = 1$ soft weight. See e.g. \cite{SoftBootstrapSUSY,FermionSoft1,FermionSoft2} for discussions on the VA scattering amplitudes, and \cite{IvanovKapustnikov,IvanovKapustnikov2} for further details on non-linear SUSY.}. This is a completely general statement if each fermion is to have a canonical Weyl kinetic term and illustrates the power of this algebraic analysis. The exceptional EFTs have the maximal possible soft scaling for a given derivative power counting and therefore standout in the space of all EFTs.

However, this algebraic approach is by no means specific to theories with linearly realised Poincar\'{e} symmetries. In this paper we adapt our approach to classify supersymmetric theories i.e. we replace the linear Poincar\'{e} symmetries assumed in \cite{RSW} with those of $\mathcal{N}=1$ supersymmetry (SUSY). From now on we refer to \cite{RSW} as part I and the present paper as part II. The general question we wish to tackle is: which Lie-superalgebras can be non-linearly realised on irreducible supermultiplets with canonical propagators and interactions at weak coupling? Given the prominence of SUSY in both particle physics and cosmological model building, an exhaustive classification in this regard would prove very useful. Recently, this study has been initiated at the level of soft scattering amplitudes \cite{SoftBootstrapSUSY} and our aim in this paper is to present a complementary, and extended, analysis at the level of Lie-superalgebras.  

We will demonstrate that this classification can be achieved by employing a neat generalisation of the distinction between essential and inessential Goldstones used in part I. There the essential Goldstones are the ones which are necessary to realise all symmetries at low energies while the inessential ones can be eliminated by inverse Higgs constraints (they could be a very important part of any (partial) UV completion \cite{UV-InverseHiggs}, however). It is the commutator between space-time translations and non-linear generators which distinguishes between the two: if a non-linear generator commutes with translations into another non-linear generator, its corresponding Goldstone is inessential and can eliminated by inverse Higgs constraints\footnote{These inessential Goldstones are always massive and can therefore be integrated out of the path integral for processes with energies below their mass. This is another way of seeing that they cannot play an essential part in the low energy realisation.}. In this paper we will make use of superspace translations to provide a further way of distinguishing between inessentials and essentials in SUSY theories. As we will show, it will be possible to impose \textit{superspace inverse Higgs constraints} which relate inessentials to the SUSY-covariant derivatives of essentials. This SUSY generalisation of inverse Higgs constraints will form a central ingredient in our analysis and will be presented in detail in section \ref{SuperspaceIH}.

In that section we also show how the generators of a non-linearly realised Lie-superalgebra are related to the \textit{superspace} expansion of the essential Goldstone modes, in direct comparison to part I where we showed that the allowed generator structure is dictated by Taylor expansions. This results in \textit{superspace inverse Higgs trees} which arise from satisfying super-Jacobi identities between two copies of (super)-translations and one non-linear generator, up to the presence of linear generators. Again, these trees encode details on the massless states in the EFT, the linearly realised symmetries, and the soft weights of component fields in a given supermultiplet. Indeed, the trees also impose relations between the soft weights of the component fields, reproducing the relations derived in \cite{SoftBootstrapSUSY} using supersymmetric Ward identities\cite{Ward1,Ward2,ScatteringAmplitudes,Ward3}. This very nicely illustrates how the two independent methods are complementary and can be used to cross-check results. Given that we do not assume anything about the form of the scattering amplitudes, our results for the soft weights are valid to all orders in perturbation theory in comparison to the SUSY Ward identities. 

The existence of canonical propagators for the component fields of the essential Goldstone supermultiplets restricts the allowed generator content further. This leads to a simplification of the inverse Higgs tree and makes exhaustive classifications possible, with the only additional work requiring one to satisfy the remaining Jacobi identities. We keep section \ref{SuperspaceIH} completely general without specifying the spin of the essential Goldstones then in the subsequent sections we specialise to examples of interest: a single chiral, Maxwell or real linear superfield in sections \ref{Chiral}, \ref{Maxwell} and \ref{RealLinear} respectively\footnote{Let us emphasise that the existence of an exceptional algebra does not imply that there is a sensible low energy realisation consisting of Goldstone modes. In part I we saw that every exceptional algebra one can construct does indeed have a realisation but as we change the linear symmetries this may not be true. We will comment on this as we go along.}. For the chiral and Maxwell superfields we perform exhaustive classifications showing that exceptional EFTs can only appear at low values for the soft weights and lead to the known theories of e.g. SUSY non-linear sigma models and the VA-DBI system in the chiral case and the VA-BI system in the Maxwell case. We will show that any EFTs with soft weights enhanced with respect to these cases cannot be exceptional i.e. the symmetries must be field-independent extended shift symmetries. In the real linear case we restrict ourselves to $\sigma \leq 3$ for the real scalar lowest component field showing, for example, that this real scalar cannot be of the Special Galileon form: the supersymmetrisation of the Special Galileon algebra does not exist (we also see this in the chiral case). 

Before moving on to the main body of the paper, in the following section we will briefly review the basics of superspace and supermultiplets, primarily to fix notation. The main body of the paper (sections \ref{SuperspaceIH} - \ref{RealLinear}) follows after and we end with our concluding remarks including possible extensions of our work. In an appendix we illustrate some aspects of the coset construction for SUSY theories by deriving the Maurer-Cartan form and superspace inverse Higgs constraints for supersymmetric Galileons.

	
\section{Superspace and superfields} \label{SuperspaceReview}
Before we begin our discussion of exceptional EFTs, let us recall some basic facts about linear supersymmetry. Our conventions are the same as Wess and Bagger \cite{WessBagger}. The natural framework to describe supersymmetric theories is superspace. This allows one to construct \emph{superfields} which are manifestly covariant under supersymmetry transformations. For $\mathcal{N}=1$ superspace we extend the usual space-time, described by coordinates $x^{\mu}$ associated with translations $P_{\mu}$, with the anti-commuting Grassmann coordinates $(\theta^{\alpha}, \bar{\theta}^{\dot{\alpha}})$ associated with the fermionic generators $(Q_{\alpha}, \bar{Q}_{\dot{\alpha}})$. 
We will employ $SU(2) \times SU(2)$ notation for all indices, using the Pauli matrices $(\sigma_\mu)_{\alpha \dot \alpha}$ e.g.~$x_{\alpha \dot{\alpha}} = (\sigma^\mu)_{\alpha \dot{\alpha}} x_\mu$ and $-2x^{\mu} = (\bar{\sigma}^{\mu})^{\alpha \dot{\alpha}}x_{\alpha \dot{\alpha}}$\footnote{We remind the reader that $(\sigma^{\mu})_{\alpha \dot{\alpha}} (\bar{\sigma}_{\mu})^{\beta \dot{\beta}} = -2 \delta_{\alpha}^{\beta}\delta_{\dot{\alpha}}^{\dot{\beta}}$ which explains the factor of $2$ in the second of these expressions.}. The coordinates of superspace are then $(x^{\alpha \dot \alpha}, \theta^{\alpha}, \bar{\theta}^{\dot{\alpha}})$, while the linearly realised generators of $\mathcal{N}=1$ super-Poincar\'{e} are given by the translations $(P_{\alpha \dot \alpha}, Q_\alpha, \bar{Q}_{\dot{\alpha}})$, as well as Lorentz transformations $(M_{\alpha \beta}, \bar{M}_{\dot \alpha \dot \beta})$ subject to the non-vanishing commutator
\begin{align} \label{SUSY}
\{ Q_\alpha , \bar{Q}_{\dot{\alpha}} \} = 2 P_{\alpha \dot{\alpha}} \,,
\end{align}
of the super-Poincar\'{e} algebra. The other commutators   define the Lorentz representation of each generator. Throughout this paper we will use the following convention for commutators between a $(n/2,m/2)$ tensor $T_{\alpha_{1},\ldots \alpha_{n} \dot{\alpha}_{1},\ldots \dot{\alpha}_{m}}$ and the Lorentz generators $M_{\beta \gamma}$, $\bar{M}_{\dot{\beta}\dot{\gamma}}$:
\begin{align}
[T_{\alpha_{1}\ldots \alpha_{n} \dot{\alpha}_{1}\ldots \dot{\alpha}_{m}} , M_{\beta \gamma}] &= 2 \, n! \, i\epsilon_{\alpha_{1} (\beta} T_{\gamma) \alpha_{2} \ldots \alpha_{n} \dot{\alpha}_{1} \ldots \dot{\alpha}_{m}} \,, \notag \\
[T_{\alpha_{1}\ldots \alpha_{n} \dot{\alpha}_{1} \ldots \dot{\alpha}_{m}} , \bar{M}_{\dot{\beta} \dot{\gamma}}] &= 2 \, m! \, i \epsilon_{\dot{\alpha}_{1} (\dot{\beta}} T_{|\alpha_{1}\ldots \alpha_{n}| \dot{\gamma}) \dot{\alpha}_{2} \ldots \dot{\alpha}_{m}} \,,
\end{align}
where we have explicitly symmetrised in $(\beta, \gamma)$ or $(\dot{\beta},\dot{\gamma})$ with weight one, where necessary. In these and all following equations, the symmetrisation with weight one of groups of indices such as $\alpha_{1}, \ldots, \alpha_{n}$ will be implicit (and similarly for the dotted indices). Given that in $SU(2) \times SU(2)$ notation traces are performed with the anti-symmetric tensors $\epsilon_{\alpha \beta}$, $\epsilon_{\dot{\alpha}\dot{\beta}}$, objects which are fully symmetric are irreducible representations, e.g. the $(1,1)$ tensor $T_{\alpha_{1}\alpha_{2}\dot{\alpha}_{1}\dot{\alpha}_{2}}$ is a symmetric, traceless, rank-2 tensor. Note that when quoting and describing different algebras, we will often omit the commutators between generators and $M_{\alpha_{1}\alpha_{2}}, \bar{M}_{\dot{\alpha}_{1}\dot{\alpha}_{2}}$ but these are always implicitly understood. 
	
A general function of superspace can be expanded as a series in the Grassmann coordinates $(\theta^{\alpha}, \bar{\theta}^{\dot{\alpha}})$, which terminates at bi-quadratic order in four dimensions due to their anti-commuting nature. We have
\begin{align} \label{superspace-exp}
\Phi(x,\theta,\bar{\theta}) = \phi(x) + \theta^\alpha \chi_\alpha(x) + \bar{\theta}_{\dot{\alpha}}\bar{\xi}^{\dot{\alpha}}(x) + \ldots + \theta^2 \bar{\theta}^2 F(x),
\end{align}
where the expansion coefficients are referred to as \emph{component fields} (here indicated for a supermultiplet with a scalar field at lowest order, but taking the same form for other Lorentz representations). Passive supersymmetry transformations are translations of the Grassmann coordinates with an accompanying shift in $x^{\alpha \dot{\alpha}}$ i.e. 
\begin{align}
\theta_\alpha \rightarrow \theta_\alpha + \epsilon_\alpha, \quad \bar{\theta}_{\dot{\alpha}} \rightarrow \bar{\theta}_{\dot{\alpha}} + \bar{\epsilon}_{\dot{\alpha}}, \quad x_{\alpha \dot \alpha} \rightarrow x_{\alpha \dot \alpha} + 2 i \epsilon_\alpha \bar{\theta}_{\dot \alpha} -2 i \theta_\alpha \bar{\epsilon}_{\dot \alpha} \,,
\end{align}
and realise the supersymmetry algebra. Note that the factor of $2$ appearing in the shift of the space-time coordinates is a consequence of $SU(2) \times SU(2)$ indices. We can reinterpret the transformation of the coordinates as an active transformation on the superspace expansion components of $\Phi(x, \theta, \bar{\theta})$. The result defines the transformation law of a {superfield} and its components, which form a (generically reducible) representation. We refer the reader to \cite{Olver} if they are unfamiliar with passive vs active transformation rules and in the remainder of this paper we will always refer to active transformations.
	
Turning to dynamics, given a superfield $\Phi$ its space-time derivative $\partial_{\alpha \dot \alpha} \Phi$ is also a superfield. However, taking derivatives with respect to the Grassmann coordinates in general does not yield a superfield. Instead, it needs to be paired up with a particular space-time derivative
\begin{align}
D_\alpha &= \frac{\partial}{\partial \theta^\alpha} + i \bar{\theta}^{\dot \alpha} \partial_{\alpha \dot \alpha} \,, \qquad
\bar{D}_{\dot{\alpha}} = -\frac{\partial}{\partial \bar{\theta}^{\dot{\alpha}}} - i \theta^\alpha  \partial_{\alpha \dot{\alpha}} \, , \label{supercovder}
\end{align}
to form supercovariant derivatives $D_\alpha$ and $\bar{D}_{\dot{\alpha}}$ which satisfy $\{ D_\alpha , \bar{D}_{\dot{\alpha}} \} = -2 i \partial_{\alpha \dot \alpha}$ and \\$\{ D_\alpha , D_{\beta}\} = \{ \bar{D}_{\dot{\alpha}} , \bar{D}_{\dot{\beta}}\} = 0$. These are a crucial ingredient when building irreducible superfields; they can be used to impose covariant constraints which project onto irreducible representations. In this paper, we will consider the following irreducible superfields:
\begin{itemize}
\item The \textbf{chiral} superfield is defined by $\bar{D}_{\dot{\alpha}} \Phi = 0$. This condition reduces the field content to a complex scalar $\phi$, a Weyl fermion $\chi_\alpha$ and a complex auxiliary scalar $F$. The chiral superfield has the following superspace expansion
\begin{align}
\Phi(x,\theta,\bar \theta) = \phi(y) + \sqrt{2}\theta \chi(y) + \theta^2 F(y) \,,
\end{align} 
where $y_{\alpha \dot \alpha}= x_{\alpha \dot \alpha} -2i\theta_\alpha \bar \theta_{\dot \alpha}$. 
\item The \textbf{Maxwell} superfield is a spinor $W_\alpha$ which satisfies the conditions $D^\alpha W_\alpha + \text{c.c.} = 0$ and the chirality condition $\bar{D}_{\dot{\alpha}} W_\alpha = 0$. It contains a Weyl fermion $\chi_\alpha$,  a gauge vector $A_{\alpha \dot{\alpha}}$ and a real auxiliary scalar $D$, and has the following superspace expansion
\begin{align}
W_\alpha = \chi_\alpha(y) + i\theta_\alpha D(y) + i\theta^\beta F_{\beta\alpha}(y)  + i\theta^2 \partial_{\alpha \dot \alpha} \bar \chi^{\dot \alpha}(y) \,, \label{superspace-Maxwell}
\end{align}
where again each component is a function of $y$ due to the chirality condition. The 2-form $F_{\alpha \beta}$ is the field strength of the vector.
\item The \textbf{real linear} superfield satisfies $L = \bar{L}$, $D^2 L = \bar{D}^2 L = 0$. Its field content is a real scalar $a$, a Weyl fermion $\chi_\alpha$ and a real vector $A_{\alpha \dot{\alpha}}$ which satisfies the condition $\partial_{\alpha \dot{\alpha}} A^{\alpha \dot{\alpha}} = 0$. The latter implies that it can be seen as the Hodge dual of a 3-form field strength $H = dB$. The full expansion in superspace is
\begin{align}
L = \, & a(x) + \theta \chi(x) + \bar{\theta}\bar{\chi}(x) - \theta^\alpha \bar{\theta}^{\dot{\alpha}}A_{\alpha \dot{\alpha}}(x) -\tfrac{i}{2}\theta^2\bar{\theta}_{\dot{\alpha}}\partial^{\alpha \dot{\alpha}} \chi_\alpha(x) \nonumber \\ &+ \tfrac{i}{2}\bar{\theta}^2\theta^{\alpha}\partial_{\alpha \dot{\alpha}} \bar{\chi}^{\dot{\alpha}}(x)  + \tfrac{1}{2}\theta^2 \bar{\theta}^2\square a(x) \,.
\end{align}
\end{itemize}
When constructing algebras and exceptional EFTs, we will consider each of these cases separately. 


\section{Goldstone modes in superspace} \label{SuperspaceIH}
\subsection*{Superspace inverse Higgs constraints}
	
In order to understand non-linear realisations in superspace, it will be useful to recall what happens in ordinary space-time with Poincar\'{e} invariant field theories. We refer the reader to part I for more details \cite{RSW} but here outline the general ideas.
	
Consider a theory with the symmetry group $G$, spontaneously broken down to a sub-group $H$. This leads to the appearance of massless Goldstone modes. Each generator $G_i$ that lives in $G / H$ induces a fluctuation $\phi^i(x)$ when acting on the vacuum field configuration $\vert 0 \rangle$:
\begin{equation}
\phi^{i}(x) G_{i} \vert 0 \rangle \,.
\end{equation}
When the broken symmetries are internal, Goldstone's theorem \cite{Goldstone} tells us that each $G_i$ leads to an independent massless Goldstone mode. However, for space-time symmetry breaking there may be degeneracies between the modes $\phi^i(x)$ even when the generators $G_i$ are independent. That is, there may be non-trivial solutions to the equation \cite{LowManohar}
\begin{align}\label{eq:LowManohar}
\phi^{i}(x) G_{i} \vert 0 \rangle = 0 \,.
\end{align} 
When such non-trivial solutions exist, we may impose this equation as a constraint to consistently project out some Goldstone modes in terms of others. We refer to modes that can be projected out as \emph{inessential} Goldstone modes, and modes that cannot as \emph{essential}. 

Acting on \eqref{eq:LowManohar} with the translation operator reveals a connection to the symmetry algebra underlying the non-linear realisation of $G / H$. The translation operator acts on both the space-time dependent Goldstone modes, on which it is represented as $-i \partial_{\alpha \dot \alpha}$, as well as on the generators $G_i$. With this understanding, the application of the one-form $\tfrac{i}{2} dx^{\alpha \dot \alpha} P_{\alpha \dot \alpha}$ yields
\begin{align}
0 &= dx^{\alpha \dot \alpha} ( \partial_{\alpha \dot \alpha} \phi^i - f_{\alpha \dot \alpha j}{}^{i}  \phi^{j} ) G_i \vert 0 \rangle, \label{IHLow}
\end{align} 
with the structure constants defined by
\begin{align}
[P_{\alpha \dot \alpha},G_{i}] = if_{\alpha \dot \alpha i}{}^{j}G_{j} + {\rm linear ~ generators} \,. \label{IHCondition}
\end{align}
Projecting \eqref{IHLow} onto a particular generator, we can impose
\begin{equation}
\partial_{\alpha \dot \alpha} \phi^i - f_{\alpha \dot \alpha j}{}^{i}  \phi^{j} + \mathcal{O}(\phi^2)\label{eq:LinearIHC} = 0 \,,
\end{equation}
i.e. we can eliminate a particular Goldstone mode $\phi^i(x)$ in terms of derivatives of $\phi^j(x)$ as long as the generator $G_j$ appears in the commutator between translations and $G_i$ i.e $[P_{\alpha \dot \alpha}, G_i] \supset i f_{\alpha \dot{\alpha}i}{}^{j} G_j$. Such a constraint is called an \emph{inverse Higgs constraint} (IHC) \cite{Spacetime2}. The linear terms in these constraints follow from the above analysis for small fluctuations, while additional terms non-linear in fields and derivatives can be calculated with the coset construction for non-linear realisations \cite{Internal1,Internal2, Spacetime1,Spacetime2}\footnote{Within the coset construction one can derive other constraints which must be satisfied by the algebra if the inverse Higgs constraints are to exist \cite{KRS,McArthur}.}. 
	
We now consider how these statements carry over from ordinary four dimensional Poincar\'e space-time to $\mathcal{N} = 1$ superspace. Consider a linearly supersymmetric theory with symmetry group $G$ broken to the sub-group $H$. Supersymmetry requires that each field is accompanied by superpartners of the same mass. Since broken generators introduce massless modes, they will at the same time introduce the appropriate superpartners. In short, we must include a full superfield $\Phi^i(x, \theta, \bar{\theta})$, for each broken generator $G_i$, again with any Lorentz indices suppressed. We represent the Goldstone mode in superspace as 
\begin{equation}
\Phi^i(x, \theta, \bar{\theta}) G_i \vert 0 \rangle \,,
\end{equation} 
where $\vert 0 \rangle$ represents the supersymmetric vacuum field configuration. As before, not all Goldstone modes have to be independent. Indeed, there may be non-trivial solutions to the equation
\begin{align}\label{eq:SuperspaceDegeneracy}
\Phi^i(x, \theta, \bar{\theta}) G_i \vert 0 \rangle = 0 \, .
\end{align}
Similarly to the purely bosonic case, we can apply translations in superspace to reveal a relation to the algebra underlying the non-linear realisation. The operator $e^{-U}d e^U$ with $U = i(\frac{1}{2}x^{\alpha \dot{\alpha}} P_{\alpha \dot{\alpha}} + \theta^{\alpha}Q_\alpha + \bar{\theta}_{\dot{\alpha}} \bar{Q}^{\dot{\alpha}})$ combines the space-time and spinor derivatives in a covariant way. The exterior derivative in superspace, expressed in the supersymmetric flat space basis of \cite{WessBagger}, becomes
\begin{equation}
d =  - \tfrac{1}{2} e^{\alpha \dot{\alpha}}\partial_{\alpha \dot{\alpha}} + e^\alpha D_\alpha + e_{\dot{\alpha}} \bar{D}^{\dot{\alpha}} \, . 
\end{equation}
Acting on \eqref{eq:SuperspaceDegeneracy}, we obtain\footnote{In this expression, the supersymmetry generators $Q_\alpha$ and $\bar{Q}_{\dot{\alpha}}$ act only on the generators, not on the fields. The exterior derivative in $e^{-U} d e^U$ acts on everything to the right, including the fields, yielding a covariant expression. We also note that in our definition of $U$ the coefficient of $ix^{\alpha \dot{\alpha}}P_{\alpha \dot{\alpha}}$ is positive such that we get the usual form of the covariant derivatives in \eqref{supercovder}. See \cite{WessBagger} for more details.} 
\begin{align}\label{eq:SUSYIHC}
\left[ -\tfrac12 e^{\alpha \dot \alpha} (\partial_{\alpha \dot \alpha} \Phi^i - f_{\alpha \dot \alpha j}{}^i \Phi^j ) + e^{\alpha} (D_{\alpha} \Phi^i - f_{\alpha j}{}^i \Phi^j ) + e_{\dot \alpha} (\bar{D}^{\dot \alpha} \Phi^i - f^{\dot \alpha}{}_{j}{}^i \Phi^j )  \right] G_i \vert 0 \rangle = 0 \,,  
\end{align}
where we have used the superspace algebra
\begin{align}
[P_{\alpha \dot \alpha},G_{i}] = - i f_{\alpha \dot \alpha i}{}^{j}G_{j} + \ldots,\quad [ Q_\alpha , G_i ]_\pm = i f_{\alpha i}{}^j G_j + \ldots \quad [ \bar{Q}_{\dot{\alpha}} , G_i ]_\pm = i f_{\dot{\alpha} i}{}^j G_j + \ldots  \label{FIHCondition} 
\end{align}
with the dots indicating unbroken generators that annihilate the vacuum. The $\pm$ sign in the subscript of a bracket indicates that it is either a commutator or anti-commutator, depending on whether the two arguments are fermionic or bosonic. 
	
In complete analogy to the space-time case, we may project \eqref{eq:SUSYIHC} onto a particular generator yielding the following possibilities
\begin{align}
&\partial_{\alpha \dot \alpha} \Phi^i - f_{\alpha \dot \alpha j}{}^i \Phi^j = \mathcal{O}(\Phi^2), \quad 
&D_{\alpha} \Phi^i - f_{\alpha j}{}^i \Phi^j = \mathcal{O}(\Phi^2) , \quad
&\bar{D}^{\dot \alpha} \Phi^i - f^{\dot \alpha}{}_{j}{}^i \Phi^j = \mathcal{O}(\Phi^2) \,,
\end{align}
where again we have indicated that these constraints are valid to leading order in fields and derivatives. The non-linear completions can again be derived using the coset construction.

We now see that it is the commutators \eqref{FIHCondition} which lead to degeneracies between Goldstone modes in superspace\footnote{As in the space-time case this is a necessary condition for the constraints to exist but is not sufficient. We will discuss this further in the next sections.}: one can solve for the Goldstone superfield $\Phi^i$ as the superspace derivative of $\Phi^j$, as long as the associated generator $G_{j}$ appears in the commutator of $G_i$ and supertranslations $Q$ or $\bar{Q}$: $[Q_\alpha, G_i] \supset  f_{\alpha i}{}^{j}G_{j}$ or $[\bar{Q}_{\dot{\alpha}}, G_i] \supset  f_{\dot{\alpha} i}{}^{j}G_{j}$. These come in addition to the usual inverse Higgs constraints which rely on the commutator between generators and space-time translations as outlined above. Our strategy will be to classify supersymmetric EFTs with non-linearly realised symmetries using these constraints to reduce to single Goldstone multiplets. From now on we refer to constraints of this type as \emph{superspace inverse Higgs constraints}.

\subsection*{Superspace inverse Higgs trees}
In the previous subsection, we saw that a Goldstone mode $\Phi^j$ can be eliminated in terms of $\Phi^i$ if the corresponding generators satisfy $[Q_\alpha, G_j] \supset f_{\alpha j}{}^{i} G_{i}$ or $[\bar{Q}_{\dot{\alpha}}, G_j] \supset  f_{\dot{\alpha} j}{}^{i} G_{i}$. Of course, it may be the case that there is a third generator $G_k$ which satisfies $[Q_\alpha, G_k] \supset  f_{\alpha k}{}^{j}  G_{j}$ or $[\bar{Q}_{\dot{\alpha}}, G_k]  \supset f_{\dot{\alpha} k}{}^{j}  G_{j}$. This gives rise to a tree of non-linearly realised generators whose corresponding Goldstones are related by superspace inverse Higgs constraints. We refer to this generator structure as a superspace inverse Higgs tree. It tells us the generator content of any algebra which can be non-linearly realised on a single Goldstone supermultiplet. The inverse Higgs tree of any supermultiplet is fixed by the Jacobi identities between two copies of supertranslations and one non-linear generator.

We now assume that there is always one non-linear generator $G_0$ which satisfies $[Q, G_0] = \ldots$ and $[\bar{Q}, G_0] = \ldots$, where the $\ldots$ contain only linear generators. The generator $G_0$ then corresponds to the essential Goldstone mode $\Phi^0$ which cannot be eliminated by any inverse Higgs constraint. Under this assumption, we showed in part I that one can (by performing the appropriate basis change) always introduce a level structure to the algebra with the level of a generator fixed by how many times we must act with translations to reach $G_{0}$. This argument carries over trivially to superspace i.e. the organisation of generators into levels is always possible. However, here we have the full superspace translations and therefore different levels can be connected by any of $(P,Q,\bar{Q})$. Schematically we have
\begin{equation}
[P,G_{n}] = G_{n-1}\,, \quad [Q, G_n] = G_{n - \frac{1}{2}}\,, \quad [\bar{Q}, G_n] = G_{n - \frac{1}{2}}\,,
\end{equation}
i.e. $Q$ and $\bar{Q}$ take us from level-$n$ to $n-\tfrac{1}{2}$ while $P$ takes us to $n-1$. We therefore label each generator according to half the number of superspace inverse Higgs relations that separate it from $G_0$. This labelling works consistently due to the SUSY algebra \eqref{SUSY}.

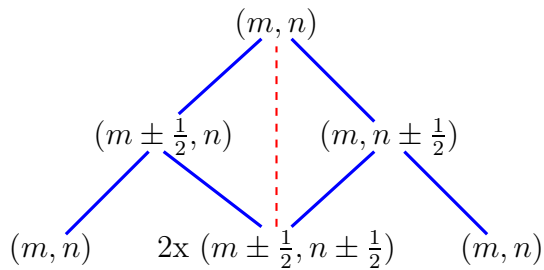
\begin{figure}
		\begin{center}
		\begin{tikzpicture}
		\draw [blue, very thick] (0+0.2,0-0.2) -- (1.5 - 0.2,-1.5 + 0.2);
		\draw [blue, very thick] (0-0.2,0-0.2) -- (-1.5 + 0.2,-1.5 + 0.3);
		\node at (0,0){$(m,n)$};
		\draw [blue, very thick](-1.5 - 0.2, -1.5 - 0.2) -- (-3 +0.2, -3 + 0.2);
		\draw [blue, very thick](-1.5+0.0,-1.5-0.2) -- (0-0.2,-3+0.3);
		\draw [blue, very thick](1.5 + 0.2, -1.5 - 0.2) -- (3 -0.2, -3 + 0.2);
		\draw [blue, very thick](1.5-0.2,-1.5-0.2) -- (0+0.2,-3+0.3);
		\node at (-1.5,-1.5) {$(m\pm \tfrac12,n)$};
		\node at (1.5,-1.5) {$(m,n\pm \tfrac12)$};
		\node at (-3,-3) {$(m,n)$};
		\node at (0,-3) {2x $(m\pm \tfrac12, n\pm \tfrac12)$};
		\draw [red, dashed, thick](0,-3+0.3) -- (0,0-0.3);
		\node at (3,-3) {$(m,n)$};
		\end{tikzpicture}
		\end{center}
		\caption{\it The non-linear generators that can be realised on a generic $(m,n)$ supermultiplet thanks to superspace inverse Higgs constraints, and their relations under superspace and space-time translations. The block blue lines heading north-west and north-east denote connections by $\bar{Q}$ and $Q$ respectively while the red dashed lines denote connections by space-time translations.}
		\label{fig:supermultiplet}
\end{figure}

Let us now see what this implies for the Goldstone modes $\Phi^i$. At level-$1/2$ in the inverse Higgs tree we find from \eqref{eq:SUSYIHC}, at linear order in fields, the following relations
\begin{align}
D_\alpha \Phi^0 = f_{\alpha \frac{1}{2}}{}^0 \Phi^{\frac{1}{2}} \,, \qquad {\bar D}_{\dot \alpha} \Phi^{0} = f_{\dot \alpha \frac{1}{2}}{}^0 \Phi^{\frac{1}{2}} \,,
\end{align}
where we allow for the essential $\Phi^0$ to be a general $(m,n)$ Lorentz representation. Clearly this implies that if the essential is bosonic (fermionic), the generators at level-$1/2$ are fermionic (bosonic). We therefore find that $(m\pm \tfrac12,n)$ and $(m,n\pm \tfrac12)$ representations can appear at this level in the tree. Including any other representations at this level would mean that the corresponding Goldstones cannot be eliminated by inverse Higgs constraints thereby increasing the number of essential modes. Moving onto level-$1$ in the tree, the inessential Goldstones corresponding to these generators can be related to the essential, via SUSY covariant derivatives, by
\begin{align}
D_\alpha D_\beta \Phi^0 = f_{\alpha \frac{1}{2}}{}^{0} f_{\beta 1}{}^{\frac{1}{2}} \Phi^1 \,, \quad
 D_\alpha {\bar D}_{\dot \beta} \Phi^0 = f_{\alpha \frac{1}{2}}{}^{0} f_{\dot \beta 1}{}^{\frac{1}{2}}  \Phi^1  \,, \notag \\ {\bar D}_{\dot \alpha} D_{\beta} \Phi^0 = f_{\dot \alpha \frac{1}{2}}{}^{0} f_{\beta 1}{}^{\frac{1}{2}} \Phi^1  \,,  \quad {\bar D}_{\dot \alpha} {\bar D}_{\dot \beta} \Phi^0 = f_{\dot \alpha \frac{1}{2}}{}^{0} f_{\dot \beta 1}{}^{\frac{1}{2}} \Phi^1 \,.
\end{align}
The derivative algebra $\{D_\alpha, D_\beta\} = 0$ implies that the LHS of the first equation is anti-symmetric and proportional to $\epsilon_{\alpha \beta}$. This imposes a constraint on the product of structure constants on the RHS. This amounts to the Jacobi identity involving the generators $(Q_\alpha, Q_\beta, G_1)$. Therefore, one finds that only the $(m,n)$ representation can be eliminated by a superspace inverse Higgs constraint using the $D^{2}$ operator and similarly for $\bar D^2$. However, the $D \bar{D}$ constraint opens up more possibilities. Indeed, there are in principle three ways to eliminate $(m \pm \tfrac12, n\pm \tfrac12)$ representations: via $D \bar D$, the opposite ordering, and by using $\partial$ i.e.
\begin{align}
\partial_{\alpha \dot \alpha} \Phi^0 = f_{\alpha \dot \alpha 1}{}^0 \Phi^1 \,.
\end{align}
The derivative algebra $\{ D_\alpha , \bar{D}_{\dot{\alpha}} \} = -2 i \partial_{\alpha \dot \alpha}$ implies that the first two of these equations adds up to the third. This requires a relationship between the structure constants, corresponding to constraints imposed by the Jacobi identity $(Q_\alpha, \bar{Q}_{\dot{\beta}}, G_1)$. There is only one of these constraints and we therefore have two copies of the four possible Lorentz representations. We have presented this superspace inverse Higgs tree in figure \ref{fig:supermultiplet} up to level-$1$. The extension to higher levels follows straightforwardly. Note that if one has EFTs with multiple essential Goldstone modes then there will be multiple inverse Higgs trees. In this paper we will work with single trees but considered multiple in part I.

The resulting set of possibilities for generators in addition to $G_{0}$ is directly related to the superspace expansion of the essential supermultiplet \eqref{superspace-exp}. This is in direct analogy to Taylor expansions in the Poincar\'{e} case. Here the superspace expansion provides a blueprint for the possible algebras that can be realised on a single essential supermultiplet. For example, the representations of modes which can be eliminated by $D^2$ and $\bar D^2$ correspond to the $\theta^2$ and $\bar \theta^2$ components. Similarly, the two copies of the $(m \pm \tfrac12, n \pm \tfrac12)$ irreps at level-$1$ are identical to the combination of the $\theta \bar \theta$ component of the superfield, as well as the $x$-expansion of its lowest $(m,n)$ component\footnote{The latter are identical to the four possibilities that we encountered in the Poincar\'{e} case at first level in those trees \cite{RSW}.}.  This pattern continues at higher levels in the tree and will be illustrated in specific cases later on. 

The inverse Higgs tree also has important implications for the transformation laws of the essential Goldstone mode. The coset construction tells us that each generator shifts its own Goldstone mode by a constant, in addition to possible field-dependent terms. Schematically we have
\begin{equation}
\delta_{G_n} \Phi^n = \epsilon^n + \ldots \,.
\end{equation}
The inverse Higgs relations then fix the field-independent part of all transformation rules. For example, since we have $D_{\alpha} \Phi^{0} = f_{\alpha \frac{1}{2}}{}^{0} \Phi^{\frac{1}{2}}$, any $G_{\frac{1}{2}}$ generator will generate a transformation rule on $\Phi^{0}$ which starts out linear in $\theta$. However, one must be careful when extending this argument to higher levels in the tree since, for example, there is no $\theta^3$ or higher component in the superspace expansion. This does not imply that there are no generators in the inverse Higgs tree connected to $G_0$ by three or more actions of $Q_\alpha$, rather Jacobi identities impose that at least one $\bar{Q}_{\dot{\alpha}}$ connection sits in between. This, in turn, implies that the inverse Higgs constraint involves at least one $\bar{D}_{\dot{\alpha}}$ on top of the three unbarred derivatives. Upon inserting $\{ D_\alpha , \bar{D}_{\dot{\alpha}} \} = -2 i \partial_{\alpha \dot \alpha}$, it is clear that the essential Goldstone mode obtains an extended shift that is (at least) linear in the space-time coordinates. This indicates that the generators are connected by a regular space-time inverse Higgs relation on top of the superspace inverse Higgs relations. Indeed, Jacobi identities demand that sequential connections by $Q$ and $\bar{Q}$ be paired up with a connection by $P$ as illustrated in figure \ref{fig:supermultiplet}.

While here we have outlined the most general superspace inverse Higgs trees that can arise, in practice we will only consider truncated versions for two reasons. The first is related to irreducibility; a generic superspace expansion forms a reducible representation of supersymmetry, and we would like to restrict ourselves to Goldstone irreps. This imposes a further restriction on the trees. The second condition follows from demanding the existence of a canonical propagator for each component within a superfield. Indeed, we demand invariance of canonical kinetic terms under the field-independent part of every non-linear transformation since this is the operator with the fewest powers of the field given that we omit tadpoles in favour of Poincar\'{e} invariant vacua\footnote{There are exceptions to this rule which rely on the presence of a dilaton and non-linear realisations of superconformal algebras which we will discuss.}. This restricts the trees even further and allows us to perform exhaustive classifications. We will comment on these additional constraints in a moment and see in practice how they are implemented in sections \ref{Chiral} - \ref{RealLinear}.

\subsection*{The coset construction in superspace}
	
Let us now outline the coset framework in superspace and connect it to our above discussion. In the standard coset construction, i.e. without SUSY, one introduces a Goldstone field for each broken generator $G_i$. Then, by computing the Maurer-Cartan form, we can read off a metric and a set of covariant derivatives which can be used to build invariant actions. We refer readers not familiar with the coset construction to the original papers \cite{Internal1,Internal2,Spacetime1} and more recent work where details are given e.g. \cite{KRS,GoonWess, Wheel}\footnote{There are also Wess-Zumino terms which as we described above play an important role in the context of extended shift symmetries. These don't follow directly from the coset construction and their derivation requires more work. See \cite{GoonWess} for a very illustrative example of finding Wess-Zumino operators for Galileons.}. As outlined above, when $[P_{\alpha \dot{\alpha}}, G_j] \supset f_{\alpha \dot{\alpha}j}{}^{i} G_i$ the Goldstone field $\phi^j(x)$ can be eliminated by an inverse Higgs constraint. In terms of the coset construction the relevant constraint is $\hat{D}_{\alpha \dot{\alpha}} \phi^i = 0$ where $\hat{D}_{\alpha \dot{\alpha}} \phi^{i}$ is the covariant derivative derived from the coset construction. This relates $\phi^j(x)$ to the space-time derivative of $\phi^i(x)$ and is simply the non-linear completion of the constraint discussed above.
		
In the SUSY case, one assigns a full Goldstone superfield to each broken generator in the coset element $\Omega$ i.e.
\begin{align}
\Omega = e^{i(\frac{1}{2}x^{\alpha \dot{\alpha}} P_{\alpha \dot{\alpha}} + \theta^{\alpha}Q_\alpha + \bar{\theta}_{\dot{\alpha}} \bar{Q}^{\dot{\alpha}})}e^{i(\Phi^0(x, \theta, \bar{\theta}) G_0)} \ldots e^{i(\Phi^N(x, \theta, \bar{\theta}) G_N)} \,,
\end{align}
where as usual we also include (super)-translations in the coset element since they act non-linearly on the superspace coordinates. From this definition of the coset element, we deduce transformation laws, a supervielbein and a set of covariant (with respect to supersymmetry and all the non-linear symmetries) derivatives. In addition to covariant space-time derivatives $\hat{D}_{\alpha \dot{\alpha}}$, we obtain modified covariant Grassmann derivatives $\hat{D}_\alpha$, $\hat{\bar{D}}_{\dot{\alpha}}$. These arise from the product of Maurer-Cartan components and the fermionic parts of the supervielbein. 
	
These covariant derivatives can now be used to impose constraints on the Goldstone superfields. The constraints separate into two classes: irreducibility constraints, which impose relations between the component fields of a particular multiplet; and superspace inverse Higgs constraints, which impose relations between multiplets i.e. in the case of a single essential are used to eliminate $\Phi^{\frac{1}{2}}, \ldots, \Phi^{N}$. We refer the reader to e.g. \cite{BaggerGalperinScalar,BaggerGalperinMaxwell,IvanovBellucciKrivonos} for more details on the superspace coset construction and to illustrate these points we present a simple example in appendix \ref{Appendix} within the context of supersymmetric Galileons which will be discussed in more detail in section \ref{Chiral}. 
	 	
\subsection*{Covariant irreducibility conditions}

As we outlined above, imposing irreducibility can constrain the structure of superspace inverse Higgs trees. Given a particular symmetry breaking pattern $G / H$, the coset construction provides a set of derivative operators $\hat{D}_\alpha$, $\hat{\bar{D}}_{\dot{\alpha}}$ that are compatible with all linear and non-linear symmetries. One should impose irreducibility in terms of these operators rather than the ordinary superspace derivatives. However, simply  imposing the naive covariantised version of the canonical constraints is not always consistent, and determining which combination of the covariant derivatives corresponds to the relevant constraint can be non-trivial   \cite{BaggerGalperinScalar,BaggerGalperinMaxwell}. We hope to clarify this issue with the following observation.

The canonical irreducibility conditions have many different symmetries. In particular, all of the symmetry algebras we classify in this paper must be realised as field transformations that preserve the irreducibility condition, and must therefore be present in the modified constraint equations for the non-linear realisation $G / H$ as well. We can make these symmetries manifest by inspecting the covariant derivatives of an extended algebra $G' / H$, which contains $G / H$ as a sub-algebra but goes up to a higher level in the superspace inverse Higgs tree. Each additional generator that we add to our algebra removes one building block for covariant constraints. Extending the algebra further and further, we eventually expect to end up with a unique building block at a particular level in the tree, which then gives rise to the covariant irreducibility condition. The correct constraint equation for $G / H$ is then also given by this covariant derivative of the extended $G' / H$, evaluated on the solution of the superspace inverse Higgs constraints. When written out in terms of the covariant derivatives of $G / H$, such a constraint can look very complicated (see \cite{BaggerGalperinMaxwell}). However, it has a simple origin in the covariant derivatives of an extended algebra. We will come across a concrete example of this in section \ref{Maxwell}.

Finally, irreducibility sometimes imposes additional constraints on the component fields. For example, the vector in the real linear multiplet satisfies $\partial_{\alpha \dot{\alpha}} A^{\alpha \dot{\alpha}} = 0$. Any symmetry transformation realised on $A_{\alpha \dot{\alpha}}$ must respect this constraint. We will examine the implications of such constraints case-by-case in the following sections.

\subsection*{Canonical propagators}

Before diving into classifying algebras and exceptional  EFTs, let us mention the second constraint on the superspace inverse Higgs tree, following from demanding canonical propagators for each component field. We recall from part I \cite{RSW} that this requirement imposes very strong constraints. For example, if the essential Goldstone is a single scalar field $\pi(x)$, all non-linear transformation rules take the form
\begin{align}
\delta_{n}\pi = s_{\mu_{1}, \ldots, \mu_{n}}x^{\mu_{1}} \ldots x^{\mu_{n}} \,,
\end{align} 
where $n$ labels the level at which the generator corresponding to the symmetric, symmetry parameter $s_{\mu_{1}, \ldots, \mu_{n}}$ appears in the scalar's tree. Note that only for $n \leq 2$ can the transformations can be augmented with field-dependent pieces \cite{RSW}. Now it is very easy to show that only the traceless part of $s$ is compatible with a canonical propagator for $\pi$ i.e. the trace part transforms the kinetic term $\pi \Box \pi$ in a way that cannot be cancelled by any other term in the Lagrangian\footnote{As we mentioned earlier, this assumes that no other operators exist at this order or below in the fields. As we explained in \cite{RSW}, the only way to violate this assumption is by adding a dilaton.}. We must therefore only include traceless generators in the scalar's tree.
	
A similar reduction in the possible generators of course occurs for fermions and vectors. In the supersymmetric setup, we require that each physical field in the supermultiplet simultaneously has a canonical propagator. Additionally, we require that the field equations for the auxiliary fields remain algebraic and contain a linear piece. In other words, we require compatibility with the following canonical superspace kinetic terms
\begin{itemize}
\item $\mathcal{L}_{free} = \int d^4 \theta \, \Phi \bar{\Phi}$ for the chiral superfield,
\item $\mathcal{L}_{free} = \int d^2 \theta \, W^\alpha W_\alpha$ for the Maxwell superfield,
\item $\mathcal{L}_{free} = \int d^4 \theta \, L^2$ for the real linear superfield.
\end{itemize}

Some of the algebras that we will encounter contain generators which induce a shift symmetry on the auxiliary fields. As auxiliary fields have algebraic field equations, the shift symmetry is broken explicitly on-shell. Therefore, the physical theory will not contain any remnant of the auxiliary field shift symmetry and we will not include the corresponding generators in our classification. Note, however, that some of the symmetry algebras we consider may be augmented by including the  auxiliary shift generators if they are automorphisms. We will discuss this point in more detail as we go along.
 
\section{Chiral supermultiplet} \label{Chiral}
\subsection*{Irreducibility condition}
We begin by illustrating the above discussion with a chiral supermultiplet $\Phi$ defined by the chirality condition $\bar{D}_{\dot{\alpha}} \Phi = 0$. In component form the chiral superfield reads
 \begin{align}
 \Phi(x,\theta,\bar \theta) = \phi(y) + \sqrt{2} \theta \chi(y) + \theta^2 F(y) \,,
 \end{align}
where $y_{\alpha \dot{\alpha}} = x_{\alpha \dot{\alpha}} -2i \theta_{\alpha} \bar \theta_{\dot{\alpha}}$ in order to satisfy the chirality condition and with $\phi$ a complex scalar, $\chi$ a Weyl spinor and $F$ an auxiliary scalar. The latter has no propagating degrees of freedom in ordinary actions (as its field equation is algebraic) but is necessary to close the supersymmetry algebra off-shell. 

Any non-linearly realised algebra must contain a $(0,0)$ complex scalar generator $G$ associated with the chiral supermultiplet $\Phi$. This follows straightforwardly from the coset construction for SUSY theories as discussed above. This generator will act non-linearly on the superfield, starting out with a constant shift and augmented with possible field-dependent pieces depending on the form of the algebra. However, the canonical superspace derivative $D_\alpha$ and its complex conjugate are not compatible with this non-linear symmetry transformation and we therefore need to make use of the modified covariant derivatives $\hat{D}_\alpha$ and $\hat{\bar{D}}_{\dot{\alpha}}$ as derived from the coset construction. By Lorentz symmetry, the most general form of the new irreducibility condition reads
\begin{align}
 T_{\dot{\alpha} \dot{\beta}}(\hat{D}\Phi, \hat{\bar{D}}\Phi,\ldots) \hat{\bar{D}}^{\dot{\beta}} \Phi  = 0 \,,
\end{align}
for some covariant operator $T_{\dot{\alpha} \dot{\beta}}$. In the following we therefore impose
\begin{equation}
\hat{\bar{D}}_{\dot{\alpha}} \Phi = 0 \,,
\end{equation}
for irreducibility regardless of the form of the non-linearly realised algebras. This clearly has important implications for the chiral field's inverse Higgs tree, since we cannot use $\hat{\bar{D}}_{\dot{\alpha}} \Phi$ to impose superspace inverse Higgs constraints. We refer the reader to \cite{BaggerGalperinScalar} for more details.

\subsection*{Superspace inverse Higgs tree}
We now turn to the chiral superfield's superspace inverse Higgs tree. We denote different levels in the tree by $n$ with half-integer levels corresponding to fermionic generators and integer levels corresponding to bosonic ones. At every level, $n$ denotes the maximum spin of an allowed generator since the essential is a scalar.

The tree starts off at $n=0$ with a complex scalar generator. Since it gives rise to an essential Goldstone, its commutator with (super)-translations can only give rise to linear generators which for now remain unconstrained. At the next level we can only add generators which live in the same representation as $\hat{D}_{\alpha} \Phi$ since $\hat{\bar{D}}_{\dot{\alpha}} \Phi$ is used to impose irreducibility. So at level $n=1/2$, we can add a single $(\frac{1}{2},0)$ Weyl fermionic generator $S_{\alpha}$, and its complex conjugate of course, with 
 \begin{align}
\{Q_\alpha, S_\beta\} = 2\epsilon_{\alpha \beta} G + \ldots,
 \end{align}
where the $\ldots$ allow for possible linear generators but not other non-linear generators. This new fermionic generator can be seen as corresponding to the component field $\chi$ in the chiral superfield, once we have imposed the relevant inverse Higgs constraint. At lowest order in fields it shifts $\Phi$ linearly in $\theta$ thereby generating a constant shift on $\chi$. Note that $[P_{\alpha \dot{\alpha}},S_{\beta}]$ and $\{\bar{Q}_{\dot{\alpha}},S_{\alpha}\}$ can give rise to linear generators but not non-linear ones. 

At level $n=1$ we can add a $(0,0)$ generator $R$, which is connected to $S_{\alpha}$ by $Q_{\alpha}$, and a $(\frac{1}{2},\frac{1}{2})$ complex vector generator\footnote{Note that here we are assuming that both scalar degrees of freedom contained in $\phi$ have identical inverse Higgs trees. This doesn't have to be the case, however. For example, we could have allowed for only a real vector generator at $n=1$ which is connected to only the real part of $\phi$. Situations like this are indeed possible. For example, we can couple a Galileon to an axion without breaking supersymmetry \cite{SoftBootstrapSUSY,SuperGals}. We consider examples of such situations in section \ref{RealLinear}. We leave an exhaustive classification for future work, but will give further comments in our conclusions.} $G_{\alpha \dot{\alpha}}$ which is connected to the essential $G$ by $P_{\alpha \dot{\alpha}}$ and to $S_{\alpha}$ by $\bar{Q}_{\dot{\alpha}}$.  The possible 2-form generator which could be connected to $S_{\beta}$ by $Q_{\alpha}$ is not consistent with Jacobi identities. In other words, the 2-form does not live in the superspace expansion of the chiral superfield. We therefore have
 \begin{align}
 [Q_{\alpha},R] = S_{\alpha} + \ldots, \quad [P_{\alpha \dot{\alpha}},G_{\beta \dot{\beta}}] = i\epsilon_{\alpha \beta} \epsilon_{\dot{\alpha} \dot{\beta}}G + \ldots, \quad [\bar{Q}_{\dot{\alpha}},G_{\beta \dot{\beta}}] = i \epsilon_{\dot{\alpha}\dot{\beta}} S_{\beta} + \ldots 
 \end{align}
The generator $R$ corresponds to a shift in $\Phi$ at quadratic order in $\theta$ and therefore generates a constant shift on the auxiliary scalar $F$.  Note that the $(Q_{\alpha},\bar{Q}_{\dot{\alpha}},G_{\beta \dot{\beta}})$ Jacobi identity requires the complex vector to have a non-vanishing commutator with both $P_{\alpha \dot{\alpha}}$ and $\bar{Q}_{\dot{\alpha}}$. This tells us that it generates a shift linear in the space-time coordinates, fitting into the Taylor expansion of the complex scalar $\phi$. At level $\theta \bar{\theta}$ we have $\partial_{\alpha \dot{\alpha}} \phi$ which indeed makes sense since this transformation is accompanied by a constant shift in $\theta \bar{\theta}$.

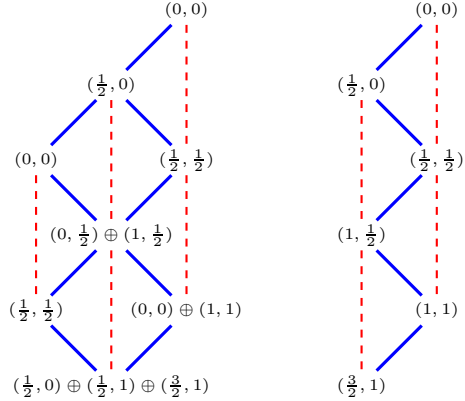
\begin{figure}
	\begin{center}
		\begin{tikzpicture}	
		\draw [blue, very thick] (0-0.2,0 - 0.2) -- (-1 + 0.2, -1 + 0.2);
		\draw [red, dashed, thick] (0, 0 - 0.2) -- (0, -2 + 0.2);
		\node at  (0,0){\tiny $(0,0)$};
		\draw [blue, very thick](-1 - 0.2, -1 - 0.2) -- (-2 + 0.2, -2 + 0.2);
		\draw [blue, very thick](-1 + 0.2, -1 -0.2) -- (0-0.2, -2 + 0.2);
		\draw [red, dashed, thick](-1, -1 - 0.2) -- (-1, -3 + 0.2);
		\node at (-1, -1) {\tiny $(\frac{1}{2},0)$};
		\draw [blue, very thick](-2 + 0.2, -2 - 0.2) -- (-1 - 0.2, -3 + 0.2);
		\draw [blue, very thick](0 - 0.2, -2 - 0.2) -- (-1 + 0.2, -3 + 0.2);
		\draw [red, dashed, thick](0, -2 - 0.2) -- (0, -4 + 0.2);
		\draw [red, dashed, thick](-2, -2 -0.2) -- (-2, -4 + 0.2);
		\node at (-2,-2){\tiny $(0,0)$};
		\node at (0, -2){\tiny $(\frac{1}{2},\frac{1}{2})$};
		\draw [blue, very thick](-1 + 0.2, -3 - 0.2) -- (0 - 0.2, -4 + 0.2);
		\draw [red, dashed, thick](-1, -3 - 0.2) -- (-1, -5 + 0.2);
		\draw [blue, very thick](-1 - 0.2, -3 - 0.2) -- (-2 + 0.2, -4 + 0.2);
		\node at (-1, -3){\tiny $(0,\frac{1}{2}) \oplus (1,\frac{1}{2})$};
		\node at (0, -4){\tiny $(0,0) \oplus (1,1)$};
		\node at (-2, -4){\tiny $(\frac{1}{2},\frac{1}{2})$};
		\draw [blue, very thick](0 - 0.2, -4 -0.2) -- (-1 +0.2, -5 + 0.2);
		\draw [blue, very thick](-2 + 0.2, -4 - 0.2) -- (-1 - 0.2, -5 + 0.2);
		\node at (-1, -5){\tiny $(\frac{1}{2},0) \oplus (\frac{1}{2},1) \oplus (\frac{3}{2},1)$};
		\end{tikzpicture}
		\hspace{0.7cm}
	\begin{tikzpicture}
		
		\draw [blue, very thick] (0-0.2,0 - 0.2) -- (-1 + 0.2, -1 + 0.2);
		\draw [red, dashed, thick] (0, 0 - 0.2) -- (0, -2 + 0.2);
		\node at (0,0){\tiny $(0,0)$};
		\draw [blue, very thick](-1 + 0.2, -1 -0.2) -- (0-0.2, -2 + 0.2);
		\draw [red, dashed, thick](-1, -1 - 0.2) -- (-1, -3 + 0.2);
		\node at (-1, -1) {\tiny $(\frac{1}{2},0)$};
		\draw [blue, very thick](0 - 0.2, -2 - 0.2) -- (-1 + 0.2, -3 + 0.2);
		\draw [red, dashed, thick](0, -2 - 0.2) -- (0, -4 + 0.2);
		\node at (0, -2){\tiny $(\frac{1}{2},\frac{1}{2})$};
		\draw [blue, very thick](-1 + 0.2, -3 - 0.2) -- (0 - 0.2, -4 + 0.2);
		\draw [red, dashed, thick](-1, -3 - 0.2) -- (-1, -5 + 0.2);
		\node at (-1, -3){\tiny $(1,\frac{1}{2})$};
		\node at (0, -4){\tiny $(1,1)$};
		\draw [blue, very thick](0 - 0.2, -4 -0.2) -- (-1 +0.2, -5 + 0.2);
		\node at (-1, -5){\tiny $(\frac{3}{2},1)$};
		\end{tikzpicture}
\caption{\it The non-linear generators that can be realised on a chiral supermultiplet (left) and the subset that is consistent with canonical propagators (right).}
\label{fig:chiraltree}
	\end{center}
\end{figure} 

The structure at higher levels follows straightforwardly with only certain representations allowed and with the connections to lower levels via (super)-translations related by Jacobi identities. On the LHS of figure \ref{fig:chiraltree} we present the tree up to level $n=5/2$. 

\subsection*{Canonical propagators}
We now consider the constraints imposed on the tree by demanding that the resulting EFT has a sensible perturbation theory: canonical propagators for physical fields augmented with weakly coupled interactions. We begin by considering the auxiliary field $F$ which in healthy theories obeys an algebraic field equation. Since the generator $R$ imposes a shift symmetry on $F$, the physical on-shell action will explicitly break this symmetry. This is telling us that we should not include this generator and indeed other generators at higher levels which are connected to $R$ by (super)-translations\footnote{As we will discuss in the next subsections, in some cases we can include the $R$ generator in a consistent manner but it is never a necessary part of the algebra. This further motivates us to omit it from the tree.} e.g. the vector at level $n=2$.

We can constrain the tree further by demanding canonical kinetic terms for $\phi$ and $\chi$ in any resulting EFT. As explained in section \ref{SuperspaceIH}, we can only add symmetric, traceless generators in the bosonic sector since these are all related to the essential complex scalar by space-time translations. For example, at $n=2$ we omit the $(0,0)$ complex generator leaving us with only the $(1,1)$ irrep. Similarly for the fermionic component field, the generators at $n=3/2$ impose a shift linear in the space-time coordinates however only the $(1,\frac{1}{2})$ generator imposes a symmetry which is consistent with the Weyl kinetic term. Again the story at higher levels is very similar to the scalar case: only a single generator is allowed and it is the one with the highest spin. Imposing these constraints on the inverse Higgs tree reduces it to the RHS of figure \ref{fig:chiraltree} with a neat zig-zag structure. We essentially have a scalar tree and a fermion tree, both with only a single branch, with the generators connected by linear SUSY. Since only a single generator appears at each level, adding a generator at say level $n=i$ requires the full tree to be present for all levels $n <i$. In the following we will denote all fermionic generators by $S$ and all bosonic ones by $G$ with the number of indices distinguishing between different levels in the tree. The complete inverse Higgs tree is therefore defined by the following (anti)-commutation relations
\begin{align}
&\{Q_\gamma, S_{\alpha_1 \ldots \alpha_{N} \dot{\alpha}_1 \ldots \dot{\alpha}_{N - 1}} \} = 2\epsilon_{\gamma \alpha_1} G_{\alpha_2 \ldots \alpha_N \dot{\alpha}_1 \ldots \dot{\alpha}_{N-1}} + \ldots \,, \nonumber \\
&[\bar{Q}_{\dot{\gamma}}, G_{\alpha_1 \ldots \alpha_N \dot{\alpha}_1 \ldots \dot{\alpha}_{N}}] = i \epsilon_{\dot{\gamma}\dot{\alpha}_1} S_{\alpha_1 \ldots \alpha_N \dot{\alpha}_2 \ldots \dot{\alpha}_N} + \ldots \,, \nonumber \\
&[P_{\gamma \dot{\gamma}}, S_{\alpha_1 \ldots \alpha_N \dot{\alpha}_1 \ldots \dot{\alpha}_{N-1}}] = i\epsilon_{\gamma \alpha_1}\epsilon_{\dot{\gamma} \dot{\alpha}_1} S_{\alpha_2 \ldots \alpha_N \dot{\alpha}_2 \ldots \dot{\alpha}_{N-1}} + \ldots \,, \nonumber \\
&[P_{\gamma \dot{\gamma}}, G_{\alpha_1 \ldots \alpha_N \dot{\alpha}_1 \ldots \dot{\alpha}_{N}}] = i\epsilon_{\gamma \alpha_1}\epsilon_{\dot{\gamma} \dot{\alpha}_1} G_{\alpha_2 \ldots \alpha_N \dot{\alpha}_2 \ldots \dot{\alpha}_{N}} + \ldots \,.
\end{align}

\subsection*{Relationship between soft weights}
Ultimately we are interested in exceptional EFTs with special IR behaviour i.e. enhanced soft limits. This tree structure already teaches us something about the relationship between the soft weights of the complex scalar and fermion component fields. For example, truncating the tree at $n=1/2$ means that there are no inverse Higgs constraints involving $P_{\alpha \dot{\alpha}}$ and therefore both the scalar and fermion have $\sigma=1$ soft behaviour since both have transformation rules which start out with a constant shift. However, if we terminate the tree at $n=1$, the scalar transformation rule induced by $G_{\alpha \dot{\alpha}}$ starts out linear in the space-time coordinates with possible field-dependent additions. The fermion can indeed transform under $G_{\alpha \dot{\alpha}}$ but the transformation rule will only contain field-dependent pieces and so will not enhance the fermion's soft behaviour. Therefore at this level the scalar will have $\sigma_{\phi}=2$ soft behaviour whereas the fermion will have $\sigma_{\chi}=1$. This clearly extends to higher levels: the soft weights can either be equal, if the tree terminates at a half-integer level, or the scalar's can be one higher if the tree terminates at an integer level:
\begin{align}
&\sigma_{\phi} = \sigma_{\chi} = n + 1/2 \,, ~~~~~~ \text{for half-integer $n$} \,, \notag \\
&\sigma_{\phi} = \sigma_{\chi} + 1 = n + 1 \,, ~~~~ \text{for integer $n$} \,. \label{softweights2}
\end{align}
This structure is dictated by linear SUSY and is exactly what was derived in \cite{SoftBootstrapSUSY} using the SUSY Ward identities. It is neat to see that the superspace inverse Higgs tree captures all this non-trivial information about the SUSY EFTs. We remind the reader that when constructing the tree we explicitly assumed that both components of the complex scalar have equivalent soft weights.

We note that when constructing theories there are possibilities of symmetry enhancements. For example, it could be that there is no realisation at a given level and by deriving invariants via the coset construction or otherwise, one finds that all operators have additional symmetries meaning that the theory really sits at a higher level. This happens with the dilaton EFT: it is not possible to write down a dilaton theory which is scale invariant but not invariant under special conformal transformations\footnote{It is however possible to have a scale invariant theory which is not fully conformal if we allow for Lorentz boosts to be spontaneously broken as in e.g. cosmology \cite{SymmetricSuperfluids}.}. In both cases we are required to build invariants operators out of diffeomorphism invariant combinations of the same effective metric $g_{\mu\nu} = e^{2\pi}\eta_{\mu\nu}$ where $\pi$ is the dilaton, which is easy to prove using the coset construction for the two symmetry breaking patterns. We will comment on symmetry enhancements where necessary in the following analysis.

\subsection*{Exceptional EFTs}
We are now in a position where we can perform an exhaustive analysis of the possible algebras which can be non-linearly realised by the single chiral superfield. We remind the reader that the superspace inverse Higgs tree is merely a necessary structure to $i$) reduce the EFT to the single chiral superfield by incorporating the necessary superspace inverse Higgs constraints and $ii$) satisfy Jacobi identities involving two copies of (super)-translations, up to the presence of linear generators. If there are no linear generators on the RHS of commutators between (super)-translations and a non-linear generator, and all commutators between a pair of non-linear generators vanish, then all Jacobi identities have been satisfied. Algebras of this type were discussed in the introduction; they lead to extended shift symmetries for each component field. However, these are very easy to construct and indeed always exist at every level in the tree. We will be primarily interested in the other type of possible algebras where transformation rules can be field-dependent, thereby leading to exceptional EFTs. 

\subsubsection*{$n=0$}
We begin with the most simple case: $n=0$ without any additional generators. Given our above discussion on soft limits, here the complex scalar will have $\sigma_{\phi} = 1$ behaviour while the fermion has $\sigma_{\chi} = 0$. The fermion can therefore be seen as a matter field whose presence is only required to maintain linear SUSY. This of course includes the case where $G$ commutes with all other generators thereby simply generating a constant shift on the complex scalar component $\phi$. This leads to supersymmetric $P(X)$ theories \cite{LehnersKhoury1}. Just as a standard $P(X)$ theory is the most simple Goldstone EFT one can write down arising  when a global $U(1)$ symmetry is spontaneously broken, this is the most simple supersymmetric Goldstone EFT (in terms of algebras and symmetries that is; the leading order operators can be somewhat complicated \cite{LehnersKhoury1}).

There are also slightly more complicated algebras at this level corresponding to supersymmetric non-linear sigma models characterised by the non-vanishing $[G,\bar{G}]$ commutator. In contrast to the purely shift symmetric case, the resulting EFTs can have field-dependent transformation rules and are therefore exceptional EFTs given our definition in this work. Indeed, the power counting in these theories is different to the naive expectation: even though we have $\sigma_{\phi} = 1$, the complex scalar can enter the action with fewer than one derivative per field. A simple example is the two-derivative action, which can be interpreted as a metric on the two-dimensional manifold spanned by the components of the scalar field. The non-linear generators $G$ and $\bar G$ imply that this manifold has two transitively acting isometries. The only such manifolds are the maximally symmetric ones, i.e.~the hyperbolic manifold $SU(1,1) / U(1)$ or the sphere $SO(3) / SO(2)$, which are well-known non-linear sigma models. We refer the reader to \cite{SoftBootstrapSUSY} and references therein for more details.

\subsubsection*{$n=1/2$}
We now consider the case where the tree terminates at $n=1/2$ with a single additional non-linear generator $S_{\alpha}$. The most general form of the commutators in addition to those of the linear realised super-Poincar\'{e} and the ones which define the Lorentz representation of the non-linear generators is
\begin{align}
&[P_{\alpha \dot{\alpha}},G] = a_{1}P_{\alpha \dot{\alpha}}, \quad [Q_{\alpha},G] = a_{2}Q_{\alpha}, \quad [\bar{Q}_{\dot{\alpha}},G] = a_{3}\bar{Q}_{\dot{\alpha}}, \nonumber \\
&[P_{\alpha \dot{\alpha}},S_{\beta}] = a_{4} \epsilon_{\alpha \beta}\bar{Q}_{\dot{\alpha}}, \quad \{Q_{\alpha},S_{\beta}\} = 2\epsilon_{\alpha \beta} G, + a_{5}M_{\alpha \beta},  \nonumber \\
&[G,\bar{G}] = a_{6}G + a_{7}\bar{G}, \quad [S_{\alpha},G] = a_{8} S_{\alpha} + a_{9}Q_{\alpha}, \quad [\bar{S}_{\dot{\alpha}},G] = a_{10} \bar{S}_{\dot{\alpha}}+ a_{11} \bar{Q}_{\dot{\alpha}},  \nonumber \\
&\{S_{\alpha},S_{\beta}\} = a_{12}M_{\alpha \beta}, \quad \{S_{\alpha},\bar{S}_{\dot{\alpha}}\} = a_{13}P_{\alpha \dot{\alpha}}.
\end{align}
Note that we didn't allow for a commutator of the form $\{\bar{Q}_{\dot{\alpha}},S_{\alpha}\} = a_{14}P_{\alpha \dot{\alpha}}$ since it can be set to zero by a change of basis. Now the Jacobi identities are very constraining, fixing all parameters to zero other than $a_{13} \equiv s$ which is unconstrained. If $s \neq 0$ we can set it to 2 by rescaling generators such that the algebra is that of $\mathcal{N} = 2$ SUSY augmented with the only inverse Higgs constraint\footnote{We keep $s \geq 0$ to ensure positivity in Hilbert space. This is a necessary requirement in any linear realisations of the symmetry algebra, but not in non-linear realisations as the currents don't integrate into well-defined charges in the quantum theory. Here we still assume the requirement of positivity in Hilbert space. This is a reasonable assumption if one anticipates that the non-linear realisations have a (partial) UV completion to a linearly realised theory, or to be a particular limit of such a theory.}. In this case the component field $\chi$ takes the Volkov-Akulov (VA) form \cite{VA}. This is an exceptional algebra by virtue of having a non-vanishing commutator between non-linear generators. On the other hand, if $s = 0$ then $S_{\alpha}$ generates a constant shift on $\chi$ as studied in \cite{InternalSUSY}. This is simply a contraction of the $s \neq 0$ algebra. In both cases $G$ generates a constant shift on the complex scalar component field $\phi$ since by Jacobi identities $G$ must commute with (super)-translations and with $\bar{G}$. We therefore have a shift symmetric complex scalar field coupled to either a VA or shift symmetric fermion field with the couplings fixed by linear SUSY. The soft weights at this level are $\sigma_{\phi}= \sigma_{\chi} = 1$. This discussion is unchanged if we add linear scalar generators\footnote{Linearly realised scalar generators commute with the Poincar\'{e} factor but can appear on the RHS of the above commutators, can form their own sub-algebra and can have non-zero commutators with non-linear generators and super-translations.}: they do not allow for additional exceptional algebras.

In terms of the low energy EFTs which can non-linearly realise these algebras, when $s=2$ it is not clear if they are independent from those which sit at level $n=1$  i.e. there could be symmetry enhancement.  It was suggested in \cite{BaggerGalperinScalar} that the symmetry is indeed enhanced to the case where the complex scalar has an additional symmetry but much more work is required to arrive at a definitive answer. However, for $s=0$ there are invariants we can write down which do not exhibit symmetry enhancement. For example, the   operator 
\begin{equation}
\int d^4 \theta\,  \partial_{\alpha \dot{\alpha}} \Phi \partial_{\beta \dot{\beta}} \Phi \partial^{\alpha \dot{\alpha}} \bar{\Phi} \partial^{\beta \dot{\beta}} \bar{\Phi} \,,
\end{equation}
for the chiral superfield $\Phi$ has a shift symmetry for its scalar and fermion components  but does not exhibit enhancement to level $n=1$.

\subsubsection*{$n=1$}
We now also include the complex vector $G_{\alpha \dot{\alpha}}$ taking us to level $n=1$. Here the soft limits are $\sigma_{\phi} = 2$ and $\sigma_{\chi} = 1$. We play the same game as before: write down the most general commutators consistent with the superspace inverse Higgs tree and impose Jacobi identities to derive the algebras which can be non-linearly realised on the chiral superfield. This is a simple generalisation of the $n=1/2$ case but since the full Ansatz for the commutators is quite involved, here we will just describe the results. As in the previous case, we allow for linear scalar generators which now turn out to be crucial in deriving exceptional algebras and EFTs. Note that in the Ansatz we do not allow for $G$ or $\bar{G}$ to appear on the RHS of a commutator between a pair of non-linear generators which correspond to inessential Goldstones ($S_{\alpha}$ and $G_{\alpha \dot{\alpha}}$). This is necessary to ensure that the relevant superspace inverse Higgs constraints exists i.e. that the inessential Goldstones appear algebraically in the relevant covariant derivatives. We refer the reader to \cite{KRS} for more details. 

Given that in all cases the bosonic generators form a sub-algebra, we can use the results of part I to fix these commutators. We refer the reader to \cite{RSW} for more details but let us briefly outline the allowed structures. As in the $n=1/2$ case, we find that the essential complex scalar cannot contain a component which transforms like a dilaton so the sub-algebra must correspond to that of the six-dimensional Poincar\'{e} group or contractions thereof. We can perform two distinct contractions thereby yielding three inequivalent algebras with their defining features the commutators between non-linear generators. The non-zero commutators which involve non-linear generators in the uncontracted six-dimensional Poincar\'{e} algebra are
\begin{align} \label{PoincareSix}
[P_{\alpha \dot{\alpha}},G_{\beta \dot{\beta}}] &=i \epsilon_{\alpha\beta}\epsilon_{\dot{\alpha}\dot{\beta}}G, \quad [G_{\alpha \dot{\alpha}},\bar{G}_{\beta \dot{\beta}}] =  -i(\epsilon_{\alpha\beta}\bar{M}_{\dot{\alpha}\dot{\beta}} + \epsilon_{\dot{\alpha}\dot{\beta}}M_{\alpha \beta}) +2\epsilon_{\alpha\beta}\epsilon_{\dot{\alpha}\dot{\beta}}M, \nonumber \\
& [\bar{G},G_{\alpha \dot{\alpha}}] = 2i P_{\alpha \dot{\alpha}}, \quad [G,M] = G, \quad [G_{\alpha \dot{\alpha}},M] =G_{\alpha \dot{\alpha}},
\end{align}
where $M$ is a real, linearly realised scalar generator. The non-linear realisation of this algebra is the two-scalar multi-DBI theory which has a neat probe brane interpretation \cite{multiDBI}. 

The obvious contraction we can do leads to the trivial algebra where all non-linear generators commute leaving only the commutators required by superspace inverse Higgs constraints (and the linearly realised bosonic sub-algebra). The low energy realisation of this algebra is that of bi-Galileons \cite{bi-gal} and can be seen as taking the small-field limit for both components of the complex scalar. However, there is also a less obvious contraction we can perform where we retain non-vanishing commutators between non-linear generators. This contraction is somewhat difficult to understand in terms of these complex generators but is simple when using the more familiar generators $P_{A}$, $M_{AB}$ where $A,B,\ldots$ are $SO(1,5)$ indices. In this case the linear scalar is $M_{45} \equiv M$ and the non-linear four-dimensional vectors are $M_{\mu 4} \equiv K_{\mu}$ and $M_{\mu 5} \equiv \hat{K}_{\mu}$, where $\mu$ is an $SO(1,3)$ index, which are related to the complex generators by 
\begin{align}
G = P_{4} + i P_{5}, \quad G_{\alpha \dot{\alpha}} = K_{\alpha \dot{\alpha}} + i \hat{K}_{\alpha \dot{\alpha}}.
\end{align}
The relevant contraction corresponds to sending $P_{5} \rightarrow \omega P_{5}, \hat{K}_{\mu} \rightarrow \omega \hat{K}_{\mu}$ and $M_{45} \rightarrow \omega M_{45}$ with $\omega \rightarrow \infty$. This contracted algebra is non-linearly realised by a DBI scalar coupled to a Galileon and can be seen as taking a small field limit for only one component of the complex scalar\footnote{This algebra also appeared in \cite{LieAlgebraicScalar2} and let us note that it is not clear if there exists a sensible realisation where both scalars have canonical kinetic terms. However, we will see in a moment that even if this theory existed, it cannot be supersymmetrised.}. If we now switch back to the complex generators, since $[P_{5},M_{\mu 5}] = 0$ we now have $[G,G_{\alpha \dot{\alpha}}] \neq 0$ in contrast to the fully uncontracted case. This will be important in what follows. We now take each of these sub-algebras in turn and ask which are consistent with linear SUSY and the required non-linear fermionic generator $S_{\alpha}$.

If the bosonic sub-algebra is given by \eqref{PoincareSix} then we find, perhaps unsurprisingly, that the most general algebra is that of six-dimensional super-Poincar\'{e}. In addition to the linearly realised super-Poincar\'{e} algebra and \eqref{PoincareSix}, the non-zero commutators are
\begin{align}
 \{Q_{\alpha}, S_{\beta}\} = 2\epsilon_{\alpha \beta}G , \quad \{S_{\alpha}, \bar{S}_{\dot{\alpha}}\} =2 P_{\alpha \dot{\alpha}}, \quad  [Q_{\alpha}, \bar{G}_{\beta \dot{\beta}}] = i\epsilon_{\alpha \beta} \bar{S}_{\dot{\beta}}, \quad  [S_{\alpha}, \bar{G}_{\beta \dot{\beta}}] = -i\epsilon_{\alpha \beta} \bar{Q}_{\dot{\beta}}.
\end{align}
In the resulting low energy realisation, the complex scalar takes the multi-DBI form while the fermion takes the VA form. This theory has been very well studied in various contexts, see e.g. \cite{BaggerGalperinScalar,RocekTseytlin}. 

If the bosonic algebra is the bi-Galileon one i.e. where the only non-vanishing commutators are those required by inverse Higgs constraints, we find that the supersymmetrisation also requires all commutators between non-linear generators to vanish. The only non-trivial commutators are therefore those required by superspace inverse Higgs constraints. This is simply a contraction of the six-dimensional Poincar\'{e} algebra and results in the six-dimensional supersymmetric Galileon algebra. Here the fermion is shift symmetric and a quartic Wess-Zumino interaction for this algebra was constructed in \cite{PhotonFarakos} (for more details see \cite{SoftBootstrapSUSY,SuperGals,InternalSUSY}). We present the coset construction for this symmetry breaking pattern in appendix \ref{Appendix}.

Turning to the final bosonic sub-algebra, we find that it is impossible to supersymmetrise the theory of a DBI scalar coupled to a Galileon. Indeed, the Jacobi identities involving $(Q_{\alpha},\bar{Q}_{\dot{\alpha}}, G_{\beta \dot{\beta}})$ and $(Q_{\alpha},S_{\beta}, G_{\gamma \dot{\gamma}})$ fix $[G,G_{\alpha \dot{\alpha}}] = 0$ which is incompatible with this partly contracted algebra. We therefore conclude that there is only a single exceptional EFT for a chiral superfield with $\sigma_{\phi} = 2$, $\sigma_{\chi} = 1$ soft limits which is the VA-DBI system which non-linearly realises the six-dimensional super-Poincar\'{e} algebra.

\subsubsection*{$n \geq 3/2$}
When $n \geq 3/2$ we find that no exceptional EFTs are possible: the only non-trivial commutators are the ones required by superspace inverse Higgs constraints and lead to extended shift symmetries for the component fields. The situation for $n=3/2$ is slightly different than for $n \geq 2$ so we will discuss these in turn but the results are qualitatively the same. 

At $n=3/2$, the bosonic sub-algebra must again be that of six-dimensional Poincar\'{e}, or contractions, since $i)$ the fermionic generators do not allow for a dilaton as one component of the chiral superfield and $ii)$ compared to $n=1$ we haven't added any additional bosonic generators. However, we very quickly establish that this sub-algebra must be the fully contracted one i.e. both components of the complex scalar must transform as Galileons as opposed to DBI scalars. 

To arrive at this conclusion we first use the $(P_{\alpha \dot{\alpha}}, P_{\beta \dot{\beta}}, S_{\gamma_{1}\gamma_{2}\dot{\gamma}})$ Jacobi identity to fix $[P_{\alpha \dot{\alpha}},S_{\beta}] = 0$ and the $(P_{\alpha \dot{\alpha}},S_{\beta}, \bar{S}_{\dot{\beta}})$ Jacobi identity to eliminate $G_{\alpha \dot{\alpha}}$ and $\bar{G}_{\alpha \dot{\alpha}}$ from the RHS of $\{S_{\alpha}, \bar{S}_{\dot{\alpha}}\}$. From the Jacobi identities involving two copies of (super)-translations and $S_{\alpha}$ we fix $G$ to commute with all (super)-translations and remove the possibility of adding Lorentz generators to the RHS of $\{Q_{\alpha}, S_{\beta}\}$. The Jacobi identities involving one (super)-translation, $G$ and either of the fermionic non-linear generators, and the $(Q_{\alpha},S_{\beta},\bar{S}_{\dot{\beta}})$ Jacobi, ensures that $G$ commutes with these fermionic generators. From the $(G,Q_{\alpha}, S_{\beta_{1}\beta_{2}\dot{\beta}})$ and $(\bar{G},Q_{\alpha}, S_{\beta_{1}\beta_{2}\dot{\beta}})$ Jacobi identities we then see that $[G, G_{\alpha \dot{\alpha}}] = [\bar{G},G_{\alpha \dot{\alpha}}] = 0$ thereby telling us that the bosonic sub-algebra must be the fully contracted one. The remaining Jacobi identities tell us that all other commutators between non-linear generators must vanish leaving us with only extended shift symmetries. We have checked that this conclusion is unaltered if we allow for linear scalars generators beyond the one in the bosonic sub-algebra. So for $\sigma_{\phi} = \sigma_{\chi} = 2$ there are no exceptional EFTs.

The cases with $n \geq 2$ are slightly more straightforward given our results in part I. There we showed that if the essential Goldstone is a complex scalar, there are no exceptional EFTs with $\sigma_{\phi} \geq 3$. That is, if we include the $(\frac{1}{2},\frac{1}{2})$ complex generator $G_{\alpha \dot{\alpha}}$ and the $(1,1)$ complex generator $G_{\beta_{1}\beta_{2}\dot{\beta}_{1}\dot{\beta}_{2}}$, all non-linear generators must commute and give rise to only extended shift symmetries. In particular, there is no complex version  of the Special Galileon, the algebra simply doesn't exist. Taking this as a starting point, we add the necessary superspace inverse Higgs commutators and use Jacobi identities to show that all non-linear generators, bosonic and fermionic, must commute amongst themselves. The calculation follows in a similar spirit to those described above and is valid for any finite $n \geq 2$.

\subsubsection*{Brief summary}
Just like in part I, we have seen that exceptional EFTs are hard to come by: there are   only a small number of non-linearly realised algebras which allow for field-dependent transformation rules on a chiral superfield. Here we summarise the main results of this section:
\begin{itemize}
\item The structure of the chiral superfield's superspace inverse Higgs tree tells us that the soft weights of the component fields are either equal or the complex scalar's can be one higher. The soft weights are fixed by the level of the inverse Higgs tree and given by \eqref{softweights2}.
\item The most simple exceptional EFTs are non-linear sigma models characterised by $[G,\bar{G}] \neq 0$. Here the scalar has a $\sigma_{\phi}=1$ soft weight whereas the fermion must have $\sigma_{\chi} = 0$. Indeed, whenever we include the generator $S_{\alpha}$, which is necessary for $\sigma_{\chi} \geq 1$, we find $[G,\bar{G}] = 0$.
\item In addition to non-linear sigma models, the only possible exceptional EFTs have $\sigma_{\chi} = 1$ and $\sigma_{\phi} = 1 ~\text{or}~ 2$. Even though an exceptional algebra exists at level $n=1/2$, we expect that there is no realisation with the corresponding properties, i.e. all EFTs one can derive will actually realise the unique $n=1$ exceptional algebra of six-dimensional super-Poincar\'{e}. The contraction of this algebra gives rise to supersymmetric Galileons.
\item All other algebras, at any other finite level in the tree, lead to field-independent extended shift symmetries. In particular, when both parts of the complex scalar have equivalent inverse Higgs trees, it is impossible to realise superconformal algebras on the single chiral superfield. We will relax the assumption of equivalent inverse Higgs trees in section \ref{RealLinear}. Furthermore, one cannot supersymmetrise the Special Galileon, at least in four dimensions. 

\item For leading values of the soft weights our results are completely compatible with the on-shell approach of\cite{SoftBootstrapSUSY}.
\end{itemize} 

\section{Maxwell supermultiplet}\label{Maxwell}
\subsection*{Irreducibility condition I}
	
We now investigate the case where the zeroth order generator in the tree is a spin-$1/2$ fermion. The essential Goldstone mode is therefore a spinor superfield $W_\alpha$. After imposing irreducibility conditions, $W_\alpha$ becomes a Maxwell superfield\footnote{There is another way to obtain an irreducible multiplet from a chiral spinor superfield $\phi_\alpha$. The inverse Higgs tree of the chiral spinor allows for a gauge symmetry parametrised by a real superfield $K =\bar{K}$: $\delta \phi_\alpha = \bar{D}^2 D_\alpha K$. After gauge fixing, the field content coincides with the real linear multiplet. As this amounts to a reordering of the symmetry algebras of the previous section, we will not consider this possibility further.}${}^{,}$ \footnote{The Maxwell multiplet is ordinarily introduced as a real superfield $V =\bar{V}$, which contains a large amount of gauge redundancy. After fixing to Wess-Zumino gauge, leaving only the ordinary gauge freedom of the vector, the superfield has the same content as $W_\alpha$. The relation between the two is $W_\alpha = -\frac{1}{4}\bar{D}^2 D_\alpha V$.}. 

The Maxwell superfield is defined by two separate irreducibility conditions, whose discussion we will split up. For the moment we only consider the chirality condition
\begin{align} 
\bar{D}_{\dot{\alpha}} W_\alpha = 0 \,, \label{irred-Maxwell1}
\end{align}	
telling us that the Maxwell superfield has the usual expansion with $y_{\alpha \dot{\alpha}} = x_{\alpha \dot{\alpha}} -2i \theta_{\alpha} \bar{\theta}_{\dot{\alpha}}$ dependent coefficients. The component fields are a spin-$1/2$ fermion at lowest order, a complex scalar and 2-form at order $\theta$, and a second fermion at order $\theta^2$. 

The correct generalisation of the chirality constraint in the presence of non-linear symmetries is the obvious covariantisation which we have discussed previously
\begin{equation}
\hat{\bar{D}}_{\dot{\alpha}} W_\alpha = 0 \, ,
\end{equation}
where the hat indicates derivatives covariantised with respect to the non-linear symmetry algebra which one can derive from the coset construction. Similar to the chiral multiplet discussed in section \ref{Chiral}, this constraint is fixed by Lorentz symmetry \cite{BaggerGalperinMaxwell} and extends to all levels in the inverse Higgs tree, with the derivative replaced by the appropriately extended one. To see this, note that we look for a $(\tfrac{1}{2}, \tfrac{1}{2})$ equation built out of covariant derivatives of $W_\alpha$. The most general such equation is proportional to $\hat{\bar{D}}_{\dot{\alpha}} W_\alpha$ i.e.
\begin{equation}
T_{\alpha \dot{\alpha} \beta \dot{\beta}} \hat{\bar{D}}^{\dot{\alpha}} W^\alpha = 0 \, .
\end{equation}
The relevant solutions to this equation will also satisfy $\hat{\bar{D}}_{\dot{\alpha}} W_\alpha = 0$.

\subsection*{Superspace inverse Higgs tree}

Starting with a chiral $(\tfrac{1}{2},0)$ fermionic generator at zeroth order, we go up in the tree using the super-translations $(Q_{\alpha}, \bar{Q}_{\dot{\alpha}})$. As before, the level $n$ of a generator is half the number of steps it takes to reach zeroth order. We will initially derive the tree's structure by \textit{only} assuming the chirality condition $\bar{D}_{\dot{\alpha}} W_\alpha = 0$ and will constrain the tree further in the next section by imposing the remaining irreducibility conditions and the existence of canonical propagators for the component fields.
	
At level $n = 1/2$, we can add a complex scalar $(0,0)$ generator and a 2-form $(1,0)$ generator which are related to zeroth order by $Q_\alpha$. Indeed, each of these irreps fit into the superspace expansion of the chiral spinor superfield. We cannot include a $(\tfrac{1}{2}, \tfrac{1}{2})$ generator at this level since the barred covariant derivative is used to impose the chirality condition. 

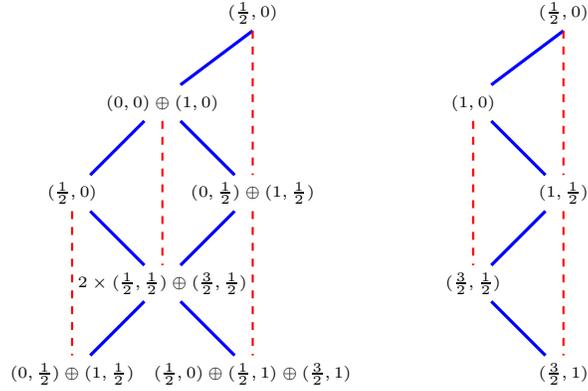
\begin{figure}[t!]
	\begin{center}
	\begin{tikzpicture}[scale = 1.2]
	\draw [blue, very thick] (0-0.0,0 - 0.2) -- (-1 + 0.2, -1 + 0.2);
	\draw [red, dashed, thick] (0, 0 - 0.2) -- (0, -2 + 0.2);
	\node at (0,0){\tiny $(\tfrac{1}{2}, 0)$};
	\draw [blue, very thick](-1 - 0.2, -1 - 0.2) -- (-2 + 0.2, -2 + 0.2);
	\draw [blue, very thick](-1 + 0.2, -1 -0.2) -- (0-0.2, -2 + 0.2);
	\draw [red, dashed, thick](-1, -1 - 0.2) -- (-1, -3 + 0.2);
	\node at (-1, -1) {\tiny $(0, 0) \oplus (1,0)$};
	\draw [blue, very thick](-2 + 0.2, -2 - 0.2) -- (-1 - 0.2, -3 + 0.2);
	\draw [blue, very thick](0 - 0.2, -2 - 0.2) -- (-1 + 0.2, -3 + 0.2);
	\draw [red, dashed, thick](0, -2 - 0.2) -- (0, -4 + 0.2);
	\draw [red, dashed, thick](-2, -2 -0.2) -- (-2, -4 + 0.2);
	\node at (-2,-2){\tiny $(\tfrac{1}{2}, 0)$};
	\node at (0, -2){\tiny $(0,\tfrac{1}{2}) \oplus (1, \tfrac{1}{2})$};
	\draw [blue, very thick](-1 + 0.2, -3 - 0.2) -- (0 - 0.2, -4 + 0.2);
	\draw [blue, very thick](-1 - 0.2, -3 - 0.2) -- (-2 + 0.2, -4 + 0.2);
	\node at (-1, -3){\tiny $2 \times (\tfrac{1}{2}, \tfrac{1}{2}) \oplus (\tfrac{3}{2},\tfrac{1}{2})$};
	\node at (0, -4){\tiny\tiny $(\tfrac{1}{2},0) \oplus (\tfrac{1}{2},1) \oplus (\tfrac{3}{2},1)$};
	\node at (-2, -4){\tiny\tiny $(0,\tfrac{1}{2}) \oplus (1,\tfrac{1}{2})$};
	\end{tikzpicture}
	\hspace{0.7cm}
	\begin{tikzpicture}[scale = 1.2]
	
	\draw [blue, very thick] (0-0.0,0 - 0.2) -- (-1 + 0.2, -1 + 0.2);
	\draw [red, dashed, thick] (0, 0 - 0.2) -- (0, -2 + 0.2);
	\node at (0,0){\tiny $(\tfrac{1}{2}, 0)$};
	\draw [blue, very thick](-1 + 0.2, -1 -0.2) -- (0-0.2, -2 + 0.2);
	\draw [red, dashed, thick](-1, -1 - 0.2) -- (-1, -3 + 0.2);
	\node at (-1, -1) {\tiny $(1,0)$};
	\draw [blue, very thick](0 - 0.2, -2 - 0.2) -- (-1 + 0.2, -3 + 0.2);
	\draw [red, dashed, thick](0, -2 - 0.2) -- (0, -4 + 0.2);
	\node at (0, -2){\tiny $(1, \tfrac{1}{2})$};
	\draw [blue, very thick](-1 + 0.2, -3 - 0.2) -- (0 - 0.2, -4 + 0.2);
	\node at (-1, -3){\tiny $(\tfrac{3}{2},\tfrac{1}{2})$};
	\node at (0, -4){\tiny $(\tfrac{3}{2},1)$};
	\end{tikzpicture}
	\caption{\it The non-linear generators that can be realised on the chiral spinor (left) and the subset that is consistent with canonical propagators and all irreducibility conditions (right).}
	\label{fig:Maxwell}
	\end{center}
	\end{figure}

At $n = 1$, only a single $(\tfrac{1}{2},0)$ generator can be connected to the $n = 1/2$ generators by $Q_{\alpha}$ even though there are two generators at that level. Indeed, Jacobi identities impose that a single spinor is connected to both $n = 1/2$ generators. Using $\bar{Q}_{\dot{\alpha}}$ to connect to $n = 1/2$, we can include $(0,\tfrac{1}{2})$ and $(1,\tfrac{1}{2})$ representations with Jacobi identities ensuring that they are also connected to zeroth order by space-time translations $P_{\alpha \dot{\alpha}}$. Here the presence of the $(0,\tfrac{1}{2})$ requires the $(0,0)$ at $n=1/2$ while the $(1,\tfrac{1}{2})$ requires the  $(1,0)$ at $n=1/2$. The extension to higher levels then follows straightforwardly in a similar fashion to what we have seen in previous sections with all generators fitting into the superspace expansion of the chiral spinor. We present this inverse Higgs tree on the LHS of figure \ref{fig:Maxwell}.
	
\subsection*{Irreducibility condition II}
 
We now turn to the remaining irreducibility condition
\begin{align}
 D^\alpha W_\alpha + \bar{D}_{\dot{\alpha}} \bar{W}^{\dot{\alpha}}  = 0 \,, \label{irred-Maxwell2}
 \end{align}
which imposes another set of constraints on the different components of this multiplet. Firstly, it reduces the complex scalar to only contain an imaginary part. Secondly, the 2-form at level $\theta$ is subject to a Bianchi identity and hence should be read as the field strength $F_{\alpha_{1} \alpha_{2}}$ of a $U(1)$ gauge vector $A_{\alpha \dot{\alpha}}$. Finally, the fermion at level $\theta^2$ becomes the derivative of the fermion at the lowest level. As a superspace expansion we therefore have
\begin{align}
W_\alpha = \chi_\alpha(y) + i\theta_\alpha D(y) + i\theta^\beta F_{\beta\alpha}(y)  + i\theta^2 \partial_{\alpha \dot \alpha} \bar \chi^{\dot \alpha}(y) \,, 
\end{align}
with a propagating fermion, the vector field strength and the real auxiliary scalar $D$. 

The covariant generalisation of the second irreducibility condition in \eqref{irred-Maxwell2} is harder to construct. We now very briefly review  \cite{BaggerGalperinMaxwell} and offer some new perspective on the uniqueness of the constraint equation found in their paper (following the general discussion in section \ref{SuperspaceIH}). Consider the anti-commutation relation $\{S_\alpha, \bar{S}_{\dot{\alpha}}\} = 2 P_{\alpha \dot{\beta}}$, with no other non-linearly realised generators. For this algebra, the naive covariantisation (placing hats on derivatives) has only the solution $W_\alpha = 0$. The correct generalisation, unique to fifth order in the fields, is given by
\begin{align}\label{eq:BaggerGalperinConstraint}
\hat{D}^{\alpha} W_{\alpha} - \frac{1}{2}\hat{D}^{\gamma} W_{\gamma}\hat{D}_{(\alpha} W_{\beta)}\hat{D}^{(\alpha} W^{\beta)} + c.c. + \ldots = 0 \,,
\end{align}
which involves both the real and imaginary parts of the trace $D^\alpha W_\alpha$ and the symmetric part of the same tensor. As explained in section \ref{SuperspaceIH} (and alluded to in \cite{BaggerGalperinMaxwell}), the origin of this peculiar constraint equation lies in its hidden covariances. To clarify its   form, we have to extend the non-linear symmetry algebra to the next level in the inverse Higgs tree discussed in the previous subsection, i.e.~include the   generators at level-$1/2$ and impose the corresponding superspace inverse Higgs constraints. 

Including only the $(0,0)$ generator $a$ at this level, we must impose the superspace inverse Higgs constraint
\begin{equation}
\hat{D}^{\alpha} W_{\alpha} - \bar{\hat{D}}_{\dot{\alpha}} \bar{W}^{\dot{\alpha}} = 0 \, .
\end{equation}
This combination of covariant derivatives cannot appear in the irreducibility condition. Combining this with the observation that the constraint \eqref{eq:BaggerGalperinConstraint} only has odd terms in the superfield, the possible combinations that one can write down in terms of the real trace and the symmetric part\footnote{One might (correctly) expect that extending the algebra to also include the representation $(1,0)$ makes this even easier, but we will find in the next subsection that this is not always possible.} are very limited. Lorentz invariance dictates that all such terms are proportional to the real trace of the covariant derivative. This implies that imposing  
\begin{align} \hat D^\alpha W_\alpha + c.c. = 0 \,, \label{constraint-Maxwell}
\end{align}
with respect to the extended algebra (including the $(0,0)$ generator) is the correct covariant irreducibility condition. As a non-trivial check of this, we have calculated that the following expression coincides with the constraints of \cite{BaggerGalperinMaxwell}
\begin{align}
(\hat{D}^{\alpha} W_{\alpha} + \hat{\bar{D}}_{\dot{\alpha}} \bar{W}^ {\dot{\alpha}})(1 + \hat{D}_{(\alpha} W_{\beta)}\hat{D}^{(\alpha} W^{\beta)} + c.c.) = 0 \, ,
\end{align}
where the hats now indicate derivatives covariantised with respect to the extended algebra including $a$. We therefore conclude that    the complicated equation of \cite{BaggerGalperinMaxwell} has its origin in a simple constraint equation of a larger symmetry algebra. 

\subsection*{Canonical propagators}

We now consider the implications on the superspace inverse tree of the second irreducibility condition \eqref{irred-Maxwell2} and the presence of canonical propagators in any resulting realisation. The combination of both of these requirements implies that the complex scalar generator at $n = 1/2$ must be omitted (its real part due to the irreducibility condition and its imaginary part to ensure that this scalar remains auxiliary). We must therefore also omit any generators which relied on its presence e.g. the $(\tfrac{1}{2},0)$ and $(0,\tfrac{1}{2})$ at $n=1$.

Continuing to higher levels we again find that only a single irrep is allowed at each level which is the one with the maximum possible spin with the spin fixed by the level $n$. For the fermion this is what we have already seen, but it also holds for the vector as discussed in part I \cite{RSW}: the only generators which do not correspond to gauge symmetries but leave the Maxwell kinetic term invariant have one pair of anti-symmetric indices with the rest fully symmetric and traceless. This corresponds to e.g. a $(\tfrac{3}{2},\tfrac{1}{2})$ hook tensor at $n=3/2$ which generates a shift on the field strength linear in the space-time coordinates. Therefore, it must act on the vector with a transformation which is quadratic in the coordinates. The full tree has again reduced to two different space-time trees, one for the fermion and one for the vector, connected by supersymmetry transformations. The vector is represented in terms of its field strength.

 In the following we again denote the fermionic generators by $S$ and the bosonic ones by $G$. The number of indices indicate where they appear in the tree. In conclusion, the superspace inverse Higgs tree is determined by the following (anti)-commutators
\begin{align}\label{eq:MaxwellIHC}
[Q_\gamma, G_{\alpha_1 \ldots \alpha_{n+3/2} \dot{\alpha}_1 \ldots \dot{\alpha}_{n-1/2}}] &= -i \epsilon_{\gamma \alpha_{n+3/2}} S_{\alpha_1 \ldots \alpha_{n+1/2} \dot{\alpha}_1 \ldots \dot{\alpha}_{n-1/2}} + \ldots \,, \nonumber \\
\{\bar{Q}_{\dot{\gamma}}, S_{\alpha_1 \ldots \alpha_{n+1} \dot{\alpha}_1 \ldots \dot{\alpha}_{n}} \} &= - \epsilon_{\dot{\gamma}\dot{\alpha}_n} G_{\alpha_1 \ldots \alpha_{n+1} \dot{\alpha}_1 \ldots \dot{\alpha}_{n-1}} + \ldots \,, \nonumber \\
[P_{\gamma \dot{\gamma}}, G_{\alpha_1 \ldots \alpha_{n+3/2} \dot{\alpha}_1 \ldots \dot{\alpha}_{n-1/2}}] &= \tfrac{i}{2}\epsilon_{\gamma \alpha_{n+3/2}}\epsilon_{\dot{\gamma} \dot{\alpha}_{n-1/2}} G_{\alpha_1 \ldots \alpha_{n+1/2} \dot{\alpha}_1 \ldots \dot{\alpha}_{n-3/2}} + \ldots \,, \nonumber \\
[P_{\gamma \dot{\gamma}}, S_{\alpha_1 \ldots \alpha_{n+1} \dot{\alpha}_1 \ldots \dot{\alpha}_{n}}] &= \tfrac{i}{2}\epsilon_{\gamma \alpha_{n+1}}\epsilon_{\dot{\gamma} \dot{\alpha}_{n}} S_{\alpha_1 \ldots \alpha_n \dot{\alpha}_1 \ldots \dot{\alpha}_{n-1}} + \ldots \,,
\end{align}
with the ellipses indicating linearly realised generators. We remind the reader that the bosonic generators only appear at half-integer levels whereas the fermionic ones appear at integer levels. This explains the otherwise peculiar labelling of indices in these equations. This truncated version of the tree is given on the RHS of figure \ref{fig:Maxwell}. Note that the gauge symmetries of the vector $A_{\alpha \dot{\alpha}}$ are not included in the tree. This is because the Maxwell superfield contains the invariant field strength in its expansion rather than the gauge potential itself. Indeed, this is why we consider the gauge multiplet in the guise of the constrained chiral superfield $W_\alpha$ rather than the vector superfield. Crucially, this allows us to restrict to a finite number of generators thereby making the tree a useful construct.
	
\subsection*{Relationship between soft weights}
We are interested in exceptional EFTs for the Maxwell superfield which have special IR behaviour in soft amplitudes. The superspace inverse Higgs tree fixes the relationship between the soft weights of the fermionic and bosonic component fields which we denote respectively as $\sigma_{\chi}$ and $\sigma_{A}$. These are again very easy to read off from figure \ref{fig:Maxwell}. As we start with a fermion at lowest order, in this case the soft weights are either equal or the fermion is one higher: 
\begin{align}
&\sigma_\chi = \sigma_A + 1 = n + 1 ~~~~ \text{for integer $n$}, \label{MaxwellSoft1} \\
&\sigma_\chi = \sigma_A = n + \tfrac{1}{2} ~~~~~~~~~ \text{for half-integer $n$} \,, \label{MaxwellSoft2}
\end{align}
which is equivalent to the relationships derived via Ward identities and soft amplitudes in \cite{SoftBootstrapSUSY}. We remind the reader that these results are valid to all orders in perturbation theory, not just at tree level, given that we have not assumed anything about the form of the amplitudes; our analysis is based purely on symmetries.

\subsection*{Exceptional EFTs}
With the superspace inverse Higgs tree at hand, we can now classify the possible exceptional algebras. We will separate our discussion into three sections: the lowest level case $n = 0$ with no superspace inverse Higgs constraints, $n = 1/2$, and finally any finite $n \geq 1$. As it turns out, the Maxwell superfield allows for only one exceptional algebra: the non-linear realisation of $\mathcal{N} = 2$ supersymmetry by a VA fermion coupled to a BI vector described by Bagger and Galperin in \cite{BaggerGalperinMaxwell}. 
	
\subsubsection*{$n = 0$}
When $n = 0$, the only non-linearly realised generator is the spinor $S_\alpha$ and therefore the Ansatz for the commutators is very simple. Jacobi identities tell us that the only non-trivial commutator involving non-linear generators is
\begin{equation}\label{eq:extendedSUSY}
\{S_\alpha, \bar{S}_{\dot{\alpha}}\} = s P_{\alpha \dot{\alpha}} \,,
\end{equation}\
which for $s =2$ leads to $\mathcal{N} = 2$ supersymmetry when combined with the other commutators. This is an exceptional algebra and is non-linearly realised by the exceptional EFT of a VA fermion coupled to a BI vector. As is now well-known \cite{LieAlgebraicVector,SoftBootstrapSUSY}, the BI vector has a vanishing soft weight and can therefore be considered as a mater field required to maintain linear SUSY. This is in comparison to the role of the fermion in $P(X)$ theories of the chiral superfield discussed in section \ref{Chiral}. The coset construction for this case was worked out in \cite{BaggerGalperinMaxwell}. The $s=0$ case is simply a contraction of the $\mathcal{N} = 2$ algebra and is non-linearly realised by a shift symmetric fermion coupled to a gauge vector in a linearly supersymmetric manner. The transformation rules here are now field-independent.

\subsubsection*{$n =1/2$} 
At level $n =1/2$, we find the real scalar generator $a$ and the 2-form $G_{\alpha_{1} \alpha_{2}}$. The real scalar generator is projected out by the requirement of canonical propagators, but we will relax this assumption for a moment and include this automorphism generator. If we only include $a$ and omit $G_{\alpha_{1} \alpha_{2}}$, Jacobi identities tell us that the only extension of the $\mathcal{N} = 2$ algebra has
\begin{align}
&[Q_\alpha, a] = S_\alpha, \quad [S_\alpha, a] = Q_\alpha \,,
\end{align}
whereas if we include $G_{\alpha_{1} \alpha_{2}}$ as well, no exceptional algebras exist\footnote{At the purely bosonic level there is a consistent exceptional algebra where the 2-form generator commutes with itself, into itself, just like the Lorentz generators. However, this algebra is not compatible with the Bianchi identity for the field strength and so cannot be realised on the gauge vector. One can see this by working out the transformation rules using the coset construction, or by reintroducing the gauge symmetries in the algebra computation as an infinite set of generators, realised on an essential vector, then checking closure of the algebra. See also \cite{LieAlgebraicVector}.}. That is, in the presence of $G_{\alpha_{1} \alpha_{2}}$, the only non-trivial commutators are those required by superspace inverse Higgs constraints i.e.
\begin{equation}
\{Q_\alpha, a\} = S_\alpha, \quad [Q_\alpha, G_{\beta_1 \beta_2}] = \epsilon_{\alpha \beta_1} S_{\beta_2}.
\end{equation}
Here the field strength transforms with a constant shift under the 2-form parameter and therefore the vector has a Galileon type symmetry: a shift linear the space-time coordinates without field dependence. Interestingly, unlike for the scalar Galileon, there are no self-interactions for this Galileon gauge vector which do not introduce additional degrees of freedom \cite{VectorNoGo}.

\subsubsection*{$n \geq 1$}
We will now proceed further in the inverse Higgs tree, to level $n = 1$ and beyond. We make use of the superspace inverse Higgs relations \eqref{eq:MaxwellIHC} and write down a general Ansatz for the remaining (anti)-commutators. Again the answer is very long and complicated so to keep things readable we will outline how we did the calculation. 
	
As we have done for the chiral superfield, we will start with just the bosonic sub-algebra which is spanned by the Poincar\'e generators and the non-linear generators $G_{\alpha_1 \ldots \alpha_{n+3/2} \dot{\alpha}_1 \ldots \dot{\alpha}_{n-1/2}}$. For $n=1$ we have already seen that the bosonic sub-algebra must be trivial but there are possible exceptional structures at higher levels. In \cite{LieAlgebraicVector} it was shown that any vector symmetry of the form $\delta A_{\alpha_{1} \dot{\alpha}} =  b_{\alpha_{1}}{}^{\alpha_{2}} x_{\alpha_{2}\dot{\alpha}}$ cannot be augmented with field-dependent pieces in the presence of the $U(1)$ gauge symmetry. Since this symmetry therefore only generates a constant shift on the field strength we will take $[P_{\gamma \dot{\gamma}}, G_{\beta_1 \beta_2}] = 0$ as a starting point. Jacobi identities then tell us that the commutators between translations and any non-linear bosonic generator are fixed by the inverse Higgs relations i.e. the third equation in \eqref{eq:MaxwellIHC} with ellipses equal to zero, up to a basis changes. 
	
Following the general recipe outlined in part I, we now inspect the Jacobi identities involving one translation and two bosonic non-linear generators : $(P, G^n, \bar{G}^m)$ and $(P, G^n, G^m)$ where again $m,n$ are half-integer. The former implies that the commutator $[G^m, \bar{G}^n] = 0$ for any $m$ and $n$ while the latter reduces the commutators schematically to
\begin{align}
[G^{z_{b}}, G^{z_{b}}] = c M, \quad [G^{z_{b}}, G^{z_{b}-1}] = c P,
\end{align}
where $z_{b}$ indicates the finite level at which the bosonic part of the tree terminates, $M$ and $P$ refer to Lorentz generators and space-time translations respectively, and $c$ is an unconstrained coefficient. These structures are very familiar from part I \cite{RSW}, for example the DBI algebra has precisely this structure. Note that Jacobi identities also allow for the 2-form generator $G_{\alpha \dot{\alpha}}$ to appear on the RHS of the first of these commutators, however its presence would spoil the inverse Higgs constraints since they would no longer be algebraic in the relevant inessential Goldstones. We encountered a similar scenario in section \ref{Chiral}. We now consider the Jacobi identity involving three non-linear generators $(G^{z_{b}}, G^{z_{b} - 1}, \bar{G}^n)$ which fixes $c=0$ since for $n>1$ there is always at least one bosonic generator which does not commute with translations due to the inverse Higgs relations. The only non-trivial commutators involving non-linear generators in the bosonic sub-algebra are therefore those required by inverse Higgs.
	
We now include the fermionic generators with the superspace inverse Higgs relations \eqref{eq:MaxwellIHC}. It is easy to see that the Jacobi identities involving two (super)-translations and one non-linear generator ensure that the ellipses in these commutators vanish i.e. we cannot include linearly realised generators on the RHS. We also see that other commutators between (super)-translations and fermionic generators, which are not required by the superspace inverse Higgs constraints, i.e. $\{Q, S^n\}$ must also vanish.
	
The only other commutators we need to fix involve two non-linear generators with at least one of these being fermionic. There is a natural way to proceed through the remaining Jacobi identities, making use of the result that the bosonic sub-algebra is trivial. We begin, for example, with the $(\bar{Q}, G^n, S^m)$ Jacobi identity which contains a single non-trivial term given by
\begin{equation}
\{\bar{Q}_{\dot{\alpha}}, [G_{\alpha_1 \ldots \alpha_{n+3/2} \dot{\alpha}_1 \ldots \dot{\alpha}_{n-1/2}}, S_{\beta_{1} \ldots \beta_{m+1}\dot{\beta}_{1} \ldots \dot{\beta}_m}]\} = 0 \,,
\end{equation} 
which is very constraining of the RHS of $[G^n, S^m]$. Proceeding in a similar fashion with the other Jacobi identities involving one supertranslation we find that schematically we can only have 
\begin{align}
\{S^{z_f}, \bar{S}^{z_f}\} = a P, \quad \{S^{z_f}, S^{z_f}\} = b M, \quad \{S^{z_f}, S^{z_{f} - 1}\} = b P \,,
\end{align}
where $z_f$ is the finite level at which the fermionic part of the tree terminates. Again we have also imposed the extra condition that all inessential Goldstones appear algebraically in the relevant covariant derivatives. Now we see that the Poincar\'{e} factor and the fermionic generators form a sub-algebra. Therefore, we can use our results of part I \cite{RSW} where we showed that the only exceptional algebra was that of the VA theory, i.e. only the zeroth order generator can form an exceptional algebra. This requires the tree to terminate at this level. Indeed, in the presence of any other fermionic generators no exceptional algebras are possible. Since in this part we are concentrating on $n \geq 1$ where we have at least two non-linear fermionic generators, we must now set $a=b=0$.

We have therefore proven, to arbitrarily high finite level in the inverse Higgs tree, that the only exceptional linearly supersymmetric EFT that can be realised on a single Maxwell superfield is the VA/BI theory which non-linearly realises $\mathcal{N}=2$ SUSY\cite{BaggerGalperinMaxwell} with $\sigma_{\chi} = 1$, $\sigma_{A} = 0$ soft weights.

\subsubsection*{Brief summary}
Let us very briefly summarise the main results for the Maxwell superfield:
\begin{itemize}
\item The superspace inverse Higgs tree allows us to read off the soft weights of the fermion and gauge vector of the Maxwell superfield. The results are given in equations \eqref{MaxwellSoft1} and \eqref{MaxwellSoft2}.
\item The only exceptional EFT in this case corresponds to a non-linear realisation of $\mathcal{N}=2$ SUSY and is realised by a VA fermion coupled to a BI vector. The soft weights are $\sigma_{\chi}=1$ and $\sigma_{A} = 0$. 
\item All other algebras lead to field-independent non-linear symmetries i.e. extended shift symmetries. We have shown this to all finite levels in the superspace inverse Higgs tree.
\item The covariant irreducibility constraints that have been imposed on the Maxwell supermultiplet can be understood via superspace inverse Higgs constraints in terms of algebras which live at a higher level in the tree. The constraints then take a simple form.
\end{itemize} 

\section{Real linear supermultiplet}\label{RealLinear}

\subsection*{Irreducibility conditions}

We now investigate the case where the zeroth order generator is a real scalar, having considered the complex scalar and spin-$1/2$ possibilities in the previous two sections. This choice naturally picks out the real linear superfield $L$ as the essential Goldstone mode with $L$ defined by the irreducibility constraints $L = \bar{L}$ and $D^2 L = \bar{D}^2 L = 0$. 
The real linear supermultiplet has a real scalar $a$ as its lowest component and a fermion $\chi$ at order $\theta$. To complete the supermultiplet, a second bosonic degree of freedom $A_{\alpha \dot{\alpha}}$ appears at order $\theta \bar{\theta}$ and satisfies the condition
\begin{align} \label{RLvectorcondition}
\partial_{\alpha \dot{\alpha}} A^{\alpha \dot{\alpha}} =0 \,.
\end{align}
The full expansion reads 
\begin{align}
L = \, & a(x) + \theta \chi(x) + \bar{\theta}\bar{\chi}(x) - \theta^\alpha \bar{\theta}^{\dot{\alpha}}A_{\alpha \dot{\alpha}}(x) -\tfrac{i}{2}\theta^2\bar{\theta}_{\dot{\alpha}}\partial^{\alpha \dot{\alpha}} \chi_\alpha(x) \nonumber \\ &+ \tfrac{i}{2}\bar{\theta}^2\theta^{\alpha}\partial_{\alpha \dot{\alpha}} \bar{\chi}^{\dot{\alpha}}(x)  + \tfrac{1}{2}\theta^2 \bar{\theta}^2\square a(x) \,.
\end{align}

The condition \eqref{RLvectorcondition} can be interpreted as the Bianchi identity of a 3-form field strength $H = dB = \star A$. This component therefore describes a 2-form gauge potential. It is sometimes possible (depending on the non-linear symmetries of the 2-form) to dualise the 2-form on-shell into a pseudoscalar, after which one obtains the same propagating degrees of freedom as the chiral supermultiplet: two scalars and one spin-$1/2$ fermion. Indeed, the dualisation can be performed on the entire supermultiplet at once, transforming a real linear superfield into a chiral superfield. This dualisation, however, does not imply equivalence between the real linear and chiral superfields. In particular, the real linear superfield cannot break the $U(1)$ R-symmetry of $\mathcal{N} = 1$ supersymmetry. This means that the chiral supermultiplet cannot be dualised when the $R$-symmetry is broken spontaneously.  

We will keep our discussion of the real linear inverse Higgs tree completely general. However, for exceptional algebras we will focus on those cases where the real linear multiplet describes two scalar degrees of freedom. As we will show, this amounts to centrally extending the non-linear symmetry algebra. Our main interest in the real linear multiplet is that it naturally describes algebras where the scalar degrees of freedom have inequivalent space-time inverse Higgs trees which we didn't allow for in section \ref{Chiral}. Examples of such systems are coupled Galileon-axions and, as we will show, the superconformal algebra. 

Examples of the coset construction using the real linear multiplet, including discussions of covariant constraints, can be found in \cite{BaggerGalperinLinear,RocekTseytlin}.

	
\subsection*{Superspace inverse Higgs tree}

At zeroth order in the tree we have the real $(0,0)$ generator $D$. At level-$1/2$ we can add spin-$1/2$ Weyl fermions since at this level we can use $\hat{D}_{\alpha}\Phi = 0$ and $\hat{\bar{D}}_{\dot{\alpha}}\Phi = 0$ to eliminate the level-$1/2$ inessential Goldstones. However, due to the reality of the essential we can only include one of these which we denote $S_{\alpha}$. The relevant inverse Higgs commutator is
\begin{align}
\{Q_{\alpha}, S_{\beta}\} = -\epsilon_{\alpha \beta}D
\end{align}
and the effect of this generator is to shift the superfield linearly in $\theta$ i.e. it generates a constant shift on $\chi$.

Now moving onto level-$1$, we can use the space-time derivative of the essential to eliminate an inessential at this level and the SUSY covariant derivative $\hat{\bar{D}}_{\dot{\alpha}}$ of the level-$1/2$ inessential. We cannot use the unbarred covariant derivative due to the irreducibility condition of the essential superfield. Another way of seeing this is that there is no $\theta^{2}$ term in the real linear superfield expansion. It then turns out that we can add two different real $(\tfrac{1}{2}, \tfrac{1}{2})$ vector generators at this level which we denote as $K_{\alpha \dot{\alpha}}$ and $\tilde{K}_{\alpha \dot{\alpha}}$. They are connected to the lower levels by
\begin{align}
[P_{\alpha \dot{\alpha}},K_{\beta \dot{\beta}}] =- i \epsilon_{\alpha \beta}\epsilon_{\dot{\alpha}\dot{\beta}}D\,, \quad [\bar{Q}_{\dot{\alpha}},K_{\beta \dot{\beta}}] = i \epsilon_{\dot{\alpha}\dot{\beta}}S_{\beta}\,, \quad [\bar{Q}_{\dot{\alpha}},\tilde{K}_{\beta \dot{\beta}}] =\epsilon_{\dot{\alpha}\dot{\beta}}S_{\beta}
\end{align}
where the possibility of adding linear generators is implied as always. Also, any other commutators between the vectors and supertranslations can give rise to linear generators only. The first of these vectors $K_{\alpha \dot{\alpha}}$ shifts the superfield linearly in the space-time coordinates which fits into the Taylor expansion of the lowest scalar component field, while the other vector $\tilde{K}_{\alpha \dot{\alpha}}$ generates a constant shift symmetry on the constrained vector at $\theta \bar{\theta}$ in the superspace expansion. Of course we can combine these two into a single complex vector generator where the real and imaginary parts have different connections to lower levels e.g. only the real part is connected to $D$ by $P_{\alpha \dot{\alpha}}$.

\begin{figure}[t!]
	\begin{center}
		\begin{tikzpicture}	
		\draw [blue, very thick] (0-0.2,0 - 0.2) -- (-1 + 0.2, -1 + 0.2);
		\draw [red, dashed, thick] (0, 0 - 0.2) -- (0, -2 + 0.2);
		\node at  (0,0){\tiny $(0,0)$};
		\draw [blue, very thick](-1 + 0.2, -1 -0.2) -- (0-0.2, -2 + 0.2);
		\draw [red, dashed, thick](-1, -1 - 0.2) -- (-1, -3 + 0.2);
		\node at (-1, -1) {\tiny $(\frac{1}{2},0)$};
		\draw [blue, very thick](0 - 0.2, -2 - 0.2) -- (-1 + 0.2, -3 + 0.2);
		\draw [red, dashed, thick](0, -2 - 0.2) -- (0, -4 + 0.2);
		\node at (0, -2){\tiny $(\frac{1}{2},\frac{1}{2})$};
		\draw [blue, very thick](-1 + 0.2, -3 - 0.2) -- (0 - 0.2, -4 + 0.2);
		\node at (-1, -3){\tiny $(0,\frac{1}{2}) \oplus (1,\frac{1}{2})$};
		\node at (0, -4){\tiny $(0,0) \oplus (1,1) \oplus (1,0)$};
		\end{tikzpicture}
		\hspace{0.7cm}
	\begin{tikzpicture}	
		\draw [blue, very thick] (0-0.2,0 - 0.2) -- (-1 + 0.2, -1 + 0.2);
		\draw [red, dashed, thick] (0, 0 - 0.2) -- (0, -2 + 0.2);
		\node at  (0,0){\tiny $(0,0)$};
		\draw [blue, very thick](-1 + 0.2, -1 -0.2) -- (0-0.2, -2 + 0.2);
		\draw [red, dashed, thick](-1, -1 - 0.2) -- (-1, -3 + 0.2);
		\node at (-1, -1) {\tiny $(\frac{1}{2},0)$};
		\draw [blue, very thick](0 - 0.2, -2 - 0.2) -- (-1 + 0.2, -3 + 0.2);
		\draw [red, dashed, thick](0, -2 - 0.2) -- (0, -4 + 0.2);
		\node at (0, -2){\tiny $(\frac{1}{2},\frac{1}{2})$};
		\draw [blue, very thick](-1 + 0.2, -3 - 0.2) -- (0 - 0.2, -4 + 0.2);
		\node at (-1, -3){\tiny $(1,\frac{1}{2})$};
		\node at (0, -4){\tiny $(1,1)$};
		\end{tikzpicture}
\caption{\it The non-linear generators that can be realised on a real linear supermultiplet (left) and the subset that is consistent with the presence of physical theories with canonical propagators (right). In general, the bosonic generators at non-zero levels are complex but with only the real part connected to the zeroth level by space-time translations.}
\label{fig:lineartree}
	\end{center}
\end{figure}
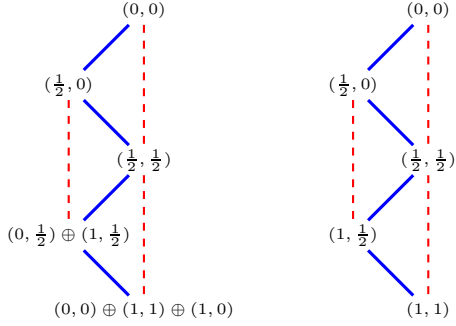

We now move to level-$3/2$ where the allowed generators must fit into the representations of the SUSY covariant derivatives of the complex vector. We find that we can add a single spin-$3/2$ generator and a single spin-$1/2$ generator. Both need to be connected to $S_{\alpha}$ by space-time translations $P_{\alpha \dot{\alpha}}$ and the full complex vector by $\bar{Q}_{\dot{\alpha}}$ or $Q_{\alpha}$. If we include this level in the tree we therefore need both the real and imaginary parts of the complex vector at level-$1$. For example, for the spin-$3/2$ generator $\psi_{\alpha_{1} \alpha_{2} \dot{\alpha}}$ the inverse Higgs commutators are
\begin{align}
[P_{\alpha \dot{\alpha}},\psi_{\beta_{1}\beta_{2} \dot{\beta}}] = i \epsilon_{\alpha \beta_{1}}\epsilon_{\dot{\alpha}\dot{\beta}}S_{\beta_{2}}\,, \quad [Q_{\alpha},\psi_{\beta_{1} \beta_{2} \dot{\beta}}] = \epsilon_{\alpha \beta_{1}}(K_{\beta_{2} \dot{\beta}} + i \tilde{K}_{\beta_{2} \dot{\beta}}).
\end{align}
Of course if we truncate the tree at level-$1$ we can include only the real or only the imaginary part of the complex vector. This pattern extends to higher levels: if we truncate the tree at a half-integer level where the highest level generators are fermionic, all bosonic generators other than the zeroth order must be complex with the real parts connected to the zeroth order generator by translations, whereas if we truncate at an integer level, the generators at the final level can also be real. This tree is presented on the LHS of figure \ref{fig:lineartree} up to level-$2$ where in comparison to the chiral case we find $(0,0)$ and $(1,1)$ generators as dictated by the Taylor expansion of the lowest component field, and a $(1,0)$ generator which lives in the Taylor expansion of the constrained vector at level $\theta \bar{\theta}$. It is connected to the imaginary part of the complex vector at level-$1$ whereas the other two generators are connected to the real part.

\subsection*{Canonical propagators}
As we have seen previously, we can constrain the form of the superspace inverse Higgs tree by demanding a canonical kinetic term $\mathcal{L}_{\text{free}} = \int d^4 \theta L^2$. Written out in component fields, this Lagrangian includes a Weyl kinetic term for the spinor $\chi_\alpha$, the Klein-Gordon kinetic term for the real scalar $\phi$, and $H^2$ for the 2-form which is dual to the constrained vector.
	
As we have seen previously, we should omit the $(\tfrac{1}{2},0)$ at level-$3/2$ and the $(0,0)$ at level-$2$. In addition, we should eliminate the $(1,0)$ at level-$2$ to be compatible with the 2-form field strength. Up to level-$2$, the tree has now reduced to the one on the RHS of figure \ref{fig:lineartree}. Note that it differs from the inverse Higgs tree of the chiral superfield in only one subtle way: the chiral case has a central extension extension $\{Q_\alpha, S_\beta\}= \ldots + i \epsilon_{\alpha \beta} Z$, with $Z$ and $D$ combining into a complex scalar generator. This simultaneously implies that $[P_{\alpha \dot{\alpha}},\tilde{K}_{\beta \dot{\beta}}] = \ldots + i \epsilon_{\alpha \beta} \epsilon_{\dot{\alpha}\dot{\beta}} Z$ due to Jacobi identities. In the end, the most general inverse Higgs tree for the real linear multiplet has
\begin{align}
&\{Q_\gamma, S_{\alpha_1 \ldots \alpha_{N} \dot{\alpha}_1 \ldots \dot{\alpha}_{N - 1}} \} = - \epsilon_{\gamma \alpha_1} G_{\alpha_2 \ldots \alpha_N \dot{\alpha}_1 \ldots \dot{\alpha}_{N-1}} + \ldots \,, \nonumber \\
&[\bar{Q}_{\dot{\gamma}}, G_{\alpha_1 \ldots \alpha_N \dot{\alpha}_1 \ldots \dot{\alpha}_{N}}] = -i \epsilon_{\dot{\gamma}\dot{\alpha}_1} S_{\alpha_1 \ldots \alpha_N \dot{\alpha}_2 \ldots \dot{\alpha}_N} + \ldots \,, \nonumber \\
&[P_{\gamma \dot{\gamma}}, S_{\alpha_1 \ldots \alpha_N \dot{\alpha}_1 \ldots \dot{\alpha}_{N-1}}] = \tfrac{1}{2}i\epsilon_{\gamma \alpha_1}\epsilon_{\dot{\gamma} \dot{\alpha}_1} S_{\alpha_2 \ldots \alpha_N \dot{\alpha}_2 \ldots \dot{\alpha}_{N-1}} + \ldots \,, \nonumber \\
&[P_{\gamma \dot{\gamma}}, G_{\alpha_1 \ldots \alpha_N \dot{\alpha}_1 \ldots \dot{\alpha}_{N}}] = \tfrac{1}{2}i\epsilon_{\gamma \alpha_1}\epsilon_{\dot{\gamma} \dot{\alpha}_1} G_{\alpha_2 \ldots \alpha_N \dot{\alpha}_2 \ldots \dot{\alpha}_{N}} + \ldots \,,
\end{align}
with the ellipses indicating linearly realised generators. In general, only the scalar generator at $n = 0$ is real.

Note that the inverse Higgs tree does not include the gauge symmetries associated to the Hodge dual 2-form either. This indicates that the 2-form gauge symmetries will never combine with the other non-linear symmetries in a non-trivial way. After imposing irreducibility conditions, again the tree only includes generators which correspond to global symmetries for the 2-form.

 \subsection*{Exceptional EFTs}
In contrast to the previous two cases, here we will not perform a general analysis. Rather we will study certain cases of interest to illustrate that our general techniques can indeed be applied to a real linear superfield. Below we consider two cases: $i)$ tree truncated at level $n=1$ with a real vector generator and $ii)$ tree truncated at level $n=2$ with the complex vector generator at $n=1$ (as required by Jacobi identities) and a real symmetric, traceless rank-2 generator (in addition to the fermionic generators in between). In the following, we only consider systems which can be dualised to the chiral superfield (or rather, those cases where the algebra does not rule out the dualisation). We leave an exhaustive classification that relaxes this assumption to future work.

\subsubsection*{$n=1$}
We begin at level $n=1$ where the non-linear generators are $(D,S_{\alpha}, K_{\alpha \dot{\alpha}})$, with $K$ Hermitian\footnote{After dualising to the chiral superfield, this is an example of an algebra where the two parts of the complex scalar zeroth order generator have different inverse Higgs trees.}. In addition to generators that define the Lorentz representation of each generator, the most general form of the commutators is
	\begin{align}
	\{Q_\alpha, \bar{Q}_{\dot{\alpha}}\} &= 2 P_{\alpha \dot{\alpha}}, \quad \{S_\alpha, \bar{S}_{\dot{\alpha}}\} = s P_{\alpha \dot{\alpha}} + a_1 K_{\alpha \dot{\alpha}}, \quad \{S_{\alpha},S_{\beta}\} = a_2 M_{\alpha \beta} \notag \\
	\{Q_\alpha, S_\beta\} &= -\epsilon_{\alpha \beta} D + a_3 M_{\alpha \beta} + i \epsilon_{\alpha \beta} M', \quad [Q_\alpha, K_{\beta \dot{\beta}}] = i \epsilon_{\alpha \beta} \bar{S}_{\dot{\beta}} + a_4 \epsilon_{\alpha \beta} \bar{Q}_{\dot{\beta}},  \notag \\
	[P_{\alpha \dot{\alpha}}, S_{\beta}] &= a_5 \epsilon_{\alpha \beta} \bar{Q}_{\dot{\alpha}}, \quad [P_{\alpha \dot{\alpha}}, D] = i a_6 P_{\alpha \dot{\alpha}}, \quad [Q_\alpha, D] = a_7 Q_\alpha, \notag \\ [K_{\alpha \dot{\alpha}}, K_{\beta \dot{\beta}}] &= a_8 \epsilon_{\alpha \beta}\bar{M}_{\dot{\alpha}\dot{\beta}} - \bar{a}_8 \epsilon_{\dot{\alpha}\dot{\beta}} M_{\alpha \beta} + i \epsilon_{\alpha \beta} \epsilon_{\dot{\alpha}\dot{\beta}} M'',  \notag \\
	[P_{\alpha \dot{\alpha}}, K_{\beta \dot{\beta}}] &= -i\epsilon_{\alpha \beta}\epsilon_{\dot{\alpha}\dot{\beta}} D + i \epsilon_{\alpha \beta} \epsilon_{\dot{\alpha}\dot{\beta}} M^{(3)} + a_9 \epsilon_{\alpha \beta} \bar{M}_{\dot{\alpha}\dot{\beta}} - \bar{a}_9 \epsilon_{\dot{\alpha}\dot{\beta}} M_{\alpha \beta}, \notag \\
	[D, S_\alpha] &= a_{10} Q_\alpha + a_{11} S_\alpha, \quad [D, K_{\alpha \dot{\alpha}}] = i a_{12} P_{\alpha \dot{\alpha}} + i a_{13} K_{\alpha \dot{\alpha}},  \notag \\
	[S_\alpha, K_{\beta \dot{\beta}}] &= a_{14} \epsilon_{\alpha \beta} \bar{Q}_{\dot{\beta}} + a_{15} \epsilon_{\alpha \beta} \bar{S}_{\dot{\beta}}.
	\end{align}
Note that we allow for the most general linear internal symmetries by introducing the scalar generators $M'$, $M''$, and $M^{(3)}$ and again we have set $\{\bar{Q}_{\dot{\alpha}}, S_{\alpha}\} = 0$ without loss of generality by a basis change. Now Jacobi identities allow for only the $M'$ linear scalar to exist and reduce the number of free parameters to two which we denote as $s$ and $m$. We have
\begin{align}\label{eq:superAdS}
\{Q_\alpha, \bar{Q}_{\dot{\alpha}}\} &= 2 P_{\alpha \dot{\alpha}}, \quad \{S_\alpha, \bar{S}_{\dot{\alpha}}\} = s P_{\alpha \dot{\alpha}} -2 m K_{\alpha \dot{\alpha}},  \notag \\
\{Q_\alpha, S_\beta\} &= -\epsilon_{\alpha \beta} D + m M_{\alpha \beta} + i \epsilon_{\alpha \beta} M', \quad [Q_\alpha, K_{\beta \dot{\beta}}] = i \epsilon_{\alpha \beta} \bar{S}_{\dot{\beta}}  \notag  \\
[P_{\alpha \dot{\alpha}}, S_{\beta}] &= -im \epsilon_{\alpha \beta} \bar{Q}_{\dot{\alpha}}, \quad [P_{\alpha \dot{\alpha}}, D] = i m P_{\alpha \dot{\alpha}}, \quad [Q_\alpha, D] = i\frac{m}{2} Q_\alpha, \notag \\ [K_{\alpha \dot{\alpha}}, K_{\beta \dot{\beta}}] &= i \frac{s}{2} \epsilon_{\alpha \beta}\bar{M}_{\dot{\alpha}\dot{\beta}} + i\frac{s}{2} \epsilon_{\dot{\alpha}\dot{\beta}} M_{\alpha \beta}, \quad [S_\alpha, K_{\beta \dot{\beta}}] = -i \frac{s}{2}\epsilon_{\alpha \beta} \bar{Q}_{\dot{\beta}},  \notag \\
[P_{\alpha \dot{\alpha}}, K_{\beta \dot{\beta}}] &= -i\epsilon_{\alpha \beta} \epsilon_{\dot{\alpha}\dot{\beta}} D + i \frac{m}{2} \epsilon_{\alpha \beta} \bar{M}_{\dot{\alpha}\dot{\beta}} + i \frac{m}{2} \epsilon_{\dot{\alpha}\dot{\beta}} M_{\alpha \beta},  \notag \\
[D, S_\alpha] &=  i\frac{m}{2} S_\alpha, \quad [D, K_{\alpha \dot{\alpha}}] = -i s P_{\alpha \dot{\alpha}} + i m K_{\alpha \dot{\alpha}}.
\end{align}
Let us now discuss these algebras in terms of $s$ and $m$. 

First of all, when $m \neq 0$ this is the $AdS_5$ superalgebra. In this case the parameter $s$ turns out to be unphysical. Indeed we can make a simple change of basis from $(K, P)$ to $(\hat{K}, P)$ where $\hat{K}_{\alpha \dot{\beta}} = K_{\alpha \dot{\beta}} - \frac{s}{2m} P_{\alpha \dot{\beta}}$ to set $s=0$. When $s \neq 0$ this basis is usually referred to as the ``AdS'' basis while with $s=0$ we have the ``conformal basis'' \cite{Equivalence}.  Therefore, the only actual parameter is the AdS radius $R = 1/m$. In terms of the bosonic sector these two bases were considered in \cite{Mapping} where it was shown that the two different realisations in terms of a single scalar degree of freedom (the vector associated to special conformal transformations is removed by an inverse Higgs constraint) are equivalent EFTs, as expected. The scalar in these theories has a vanishing soft weight  \cite{dilaton1,dilaton2,dilaton3}. As we explained in the introduction, this is compatible with our superspace inverse Higgs tree since in this case once we canonically normalise the scalar, all transformation rules become field-dependent.
	
The coset construction for this symmetry breaking pattern i.e. the $AdS_{5}$ superalgebra broken down the four-dimensional super-Poincar\'{e} algebra was studied in \cite{IvanovBellucciKrivonos, IvanovBellucciKrivonos2} (see also \cite{DeenOvrut} for a curved space generalization). The authors constructed the leading action for a supersymmetric 3-brane in $AdS_5$, utilising a real linear superfield $L$. Their Lagrangian transforms as a total derivative under a subset of the non-linear symmetries. After dualising the 2-form in $L$ to a scalar, their Lagrangian realises an additional shift symmetry that is not visible in the inverse Higgs tree. This allows for a different starting point where the essential generator is a complex scalar, but only its real part realises non-linear symmetries in addition to the constant shift symmetries. This is because there is only a real vector generator at level-$1$ and therefore only a single scalar degree of freedom can support additional transformations. This reflects the fact that the real linear superfield can be dualised to a chiral superfield. The bosonic sector is then a dilaton (which realises the conformal symmetries) coupled to an axion.
	
The flat limit of the bulk space-time corresponds to taking $m = 0$. In this case we cannot perform the aforementioned basis change and hence the second parameter $s$ distinguishes between two different algebras. The case $s = 2$ is the flat limit of the AdS superalgebra and hence corresponds to the super-Poincar\'{e} algebra in $D=5$. However, in this limit one often has symmetry enhancement to $D = 6$ super-Poincar\'e rather than $D = 5$ thanks to the dualised 2-form field which obtains a field-dependent transformation, see \cite{RocekTseytlin, BaggerGalperinScalar}. This is related to the fact that no supersymmetric scalar 3-brane exists in $D = 5$ \cite{AchucarroTownsend}. The resulting EFT is equivalent to the scalar DBI-VA system we discussed in section \ref{Chiral}.
	
A simple coset construction argument for the absence of a 3-brane in $D = 5$ is the following. Consider two actions for a complex scalar field $\phi$: the first $\mathcal{L}_6$ where both components of the complex scalar are of the DBI form and the second $\mathcal{L}_5$ where the second scalar realises only a shift symmetry rather than the full DBI transformations. We consider only operators with no more derivatives than fields. These actions differ at lowest non-trivial order in the fields. If $\mathcal{L}_5$ and $\mathcal{L}_6$ can be supersymmetrised, the leading invariants in the coset construction follow from the supervielbein and the combination of coset derivatives $\hat{D}_{\alpha} \hat{D}_{\beta \dot{\beta}} \Phi$, since all first derivatives are set to zero to impose irreducibility and inverse Higgs constraints. Therefore, the only invariant that can contribute terms with the same number of derivatives as fields is the superdeterminant of the supervielbein. Thus, the only way that both $\mathcal{L}_5$ and $\mathcal{L}_6$ can exist is if they differ by a very specific Wess-Zumino term. This is related to the existence of non-trivial cocycles in the relative cohomology of $G / H$. This allows for supersymmetry-preserving 3-branes only in\footnote{These arguments clearly generalise to different dimensions. Whenever one introduces vector generators for higher-dimensional boosts, all invariants carry more derivatives than fields. In every case, the only term that can satisfy the power counting of canonical brane action is the superdeterminant of the supervielbein. Therefore, the unique action that can correspond to a SUSY brane is determined only by the number of translation generators. This agrees with the well-known results of \cite{AchucarroTownsend} and \cite{SuperspaceGeometry}. In the latter, all possible scalar branes were classified according to relative cohomology.}$D = 6$ \cite{IvanovBellucciKrivonos, IvanovBellucciKrivonos2}.
	
Finally, we have the $m=s=0$ case which yields the $D = 5$ supersymmetric Galileon algebra. The authors of \cite{SuperGals} conjectured that this algebra has non-trivial quartic and quintic Wess-Zumino terms (in addition to the interaction constructed in \cite{PhotonFarakos}), which also realise a second shift symmetry. It is clear from our analysis that this Galileon/axion (the axion comes from dualising the 2-form) system is naturally described by a real linear superfield. We see from the algebra that when $s=m=0$ we have $\{S_{\alpha}, \bar{S}_{\dot{\alpha}}\} = 0$ and therefore the fermion is no longer of the VA type but becomes shift symmetric.
	
\subsubsection*{$n=2$}
We now consider level $n=2$ where the non-linear generators are\\ $(D,S_{\alpha}, K_{\alpha \dot{\alpha}}, \tilde{K}_{\alpha \dot{\alpha}}, \psi_{\alpha_{1}\alpha_{2}\dot{\alpha}}, G_{\alpha_{1}\alpha_{2}\dot{\alpha}_{1}\dot{\alpha}_{2}})$. As we saw above, in the presence of $\psi$ we need to include both $K$ and $\tilde{K}$ however we keep $G$ real. Rather than performing a full analysis, we ask if the lowest component of the superfield can be a Special Galileon \cite{SpecialGalileon} with a $\sigma_{\phi} =3$ soft weight and a field-dependent transformation rule. We find, thanks to our results in part I \cite{RSW}, that this is not possible. Indeed, since we are forced to include the full complex vector, after dualisation both scalar degrees of freedom must be Galileons i.e. both have a connection to a vector at level $n=1$ by space-time translations. This implies that both have a transformation rule which starts out linear in the space-time coordinates. Now we are also asking for the lowest component to be a Special Galileon. However, we have already showed in part I that we cannot couple a Special Galileon to a Galileon: there is no corresponding symmetry breaking pattern. Now since the bosonic sector is always a sub-algebra this conclusion is robust against adding the relevant fermionic generators. We therefore conclude that the lowest component of the real linear superfield cannot be of the Special Galileon form\footnote{Note that we can couple a Special Galileon to an axion but we see from the tree that this theory cannot be supersymmetrised since the presence of $\psi$ demands that the axion becomes a Galileon.}. The only remaining possibility is that a Special Galileon exists, but that this algebra is not compatible with dualisation (i.e. the central extension). This would imply that the 2-form forms an integral part of the Goldstone EFT. We leave the classification of such possibilities to future work.

\subsection*{Brief summary}

Again let us provide a brief summary of our main results with regards to the real linear superfield:
 \begin{itemize}
 \item
The superspace inverse Higgs tree becomes particularly simple after imposing both irreducibility conditions and the existence of canonical propagators, and differs from the chiral case only by having a real (instead of a complex) scalar generator at the lowest level. If we truncate the tree at a half-integer level, all bosons other than the zeroth order must be complex. However, if we truncate at an integer level, the highest generator can also be real. Moreover, the gauge symmetry of the 2-form gauge potential sitting inside the constrained vector decouple from the tree.
 \item
We have not performed an exhaustive classification, but demonstrated that the algebras up to and including $n=1$ correspond to super-AdS in $D=5$ and super-Poincar\'{e} in $D=6$. We can perform a contraction of the latter leading to a supersymmetric Galileon algebra.
\item 
At $n=2$ we have shown that the lowest order scalar cannot be a Special Galileon with a field-dependent transformation rule if we dualise the 2-form. Indeed, then the second scalar would be a Galileon which cannot be coupled to a Special Galileon \cite{RSW}. The only way out, which is an interesting avenue for future work, is to not dualise the 2-form.
 \end{itemize}

\section{Conclusions}

The IR behaviour of EFTs is strongly restricted by any non-linearly realised symmetries they might have. This is manifest in soft scattering amplitudes and has sparked the   fruitful soft bootstrap program, aiming to build theories from the bottom up using on-shell soft data. This leads to a neat classification of EFTs which is interesting from both formal and phenomenological perspectives.  

In our previous paper \cite{RSW} we have outlined a complementary   approach to classifying EFTs with special soft behaviours based on a Lie-algebraic analysis with the resulting EFTs corresponding to Poincar\'{e} invariant QFTs of Goldstone modes. In that work we classified all possible \textit{exceptional} EFTs for multi-scalar or multi-spin-$1/2$ fermion Goldstones. From the point of view of algebras and transformation rules, the exceptional EFTs non-linearly realise algebras with non-vanishing commutators between non-linear generators. These, in turn, lead to field-dependent transformation rules for the Goldstones. In terms of soft amplitudes, the special soft behaviour of these EFTs is thanks to cancellations between contact and pole Feynman diagrams. These exceptional EFTs stand out in the space of all QFTs which is why we are motivated to classify them. In the current paper, we have extended this classification to theories with a linearly realised $\mathcal{N}=1$ SUSY. 

In our algebraic approach, a key role is played by the translations of the linearly realised (super-)Poincare algebra. Commutators between non-linear generators and translations dictate whether generators give rise to internal symmetries with essential Goldstone modes (i.e.~massless excitations in the IR) or to space-time or superspace symmetries. For the latter, a number of the Goldstone modes can be inessential, i.e.~can attain a mass and therefore be integrated out of the path integral or eliminated by inverse Higgs constraints. We have extended these constraints to superspace, enabling one to reduce the number of Goldstone modes in SUSY theories in a covariant manner.

More specifically, we have shown how the triplet of translations with $\{ Q, \bar Q \} = P$ can be used to realise larger symmetry algebras without increasing the number of Goldstone modes. Starting from a specific supermultiplet, the non-linearly realised symmetries can be organised in a superspace inverse Higgs tree. This tree is fully determined by the commutators between non-linear generators and the triplet of translations. Jacobi identities restrict the spin of generators in these trees to correspond exactly to the $(x,\theta,\bar\theta)$ expansion of the original supermultiplet. This provides both a conceptually clear and calculationally simple perspective on how to build the most general algebras that can be realised. We find it useful to think of the trees as the algebraic cousin to the on-shell soft data one provides for soft bootstraps since it encodes the details of the massless states, the linearly realised symmetries and soft theorems. As an illustration of this last point, the trees allow us to read off the soft weights of the component fields of an essential Goldstone supermultiplet.

An important ingredient in order to achieve a full classification entails a trimming down of the superspace inverse Higgs trees to only those generators that give rise to symmetry transformations compatible with canonical propagators for the component fields. In the absence of a dilaton, this is a necessary requirement for the existence of a sensible EFT with a standard perturbation theory. This requirement imposes stronger constraints on the algebras than one might originally expect and in most cases reduces the trees to contain only a single generator at each level. With these highly constrained trees at hand, one can look for exceptional algebras and EFTs by imposing the remaining Jacobi identities.

We have considered the cases of a single chiral, Maxwell or real linear supermultiplet to illustrate the power of our techniques in sections \ref{Chiral}, \ref{Maxwell} and \ref{RealLinear}. The exceptional possibilities in the chiral case are limited to SUSY non-linear sigma-models, the six-dimensional super-Poincar\'{e} algebra as well as an intermediate case which we expect doesn't actually have any realisations. This super-Poincar\'{e} algebra is non-linearly realised by the scalar DBI-VA theory which couples a scalar to a fermion. In the Maxwell case, the only exceptional algebra is that of $\mathcal{N}=2$ four-dimensional super-Poincar\'{e} which is non-linearly realised by the BI-VA system and couples a gauge vector to a fermion. Remarkably, we found that there is no exceptional algebra that includes a generator that shifts the 2-form field strength of the vector component field. In contrast to all other components, the Maxwell vector can therefore not be interpreted as the Goldstone mode of some symmetry breaking pattern, in line with the conclusions of \cite{LieAlgebraicVector, SoftBootstrapSUSY}. In both of these cases our analysis is exhaustive under the assumption that in the chiral case each component of the complex scalar has the same inverse Higgs tree. 

In the real linear case, we have not performed an exhaustive classification but rather studied cases of interest. We have shown that at level $n=1$ in the real linear's inverse Higgs tree, the algebra is the $AdS_{5}$ superalgebra from which we can make two distinct contractions such that we have three different algebras. The bosonic sectors in these theories are described by a conformal Galileon (i.e. the dilaton with higher order corrections) coupled to an axion, multi-DBI or a Galileon-axion system. At level $n=2$ we have shown that the real scalar at lowest order in the superfield cannot take the Special Galileon form if we dualise the 2-form. Indeed, the allowed algebra can only give rise to field-independent transformation rules at this level meaning that we cannot supersymmetrise the Special Galileon. We found the same conclusion in section \ref{Chiral} for the chiral superfield where the bosonic sector of the theory cannot take the required form of a complex Special Galileon.

All-in-all we have seen both in part I and this paper that exceptional EFTs are rare and only appear when the soft weights of the Goldstone modes are relatively small. This further emphasises that they are very special EFTs which certainly deserve further attention. Because we only make use of algebraic methods and the theory of non-linear realisations, our statements are valid without making assumptions on the structure of interactions in the theory and for an arbitrary finite number of generators. We anticipate that reaching the same conclusions would be very difficult using amplitude methods.

Our analysis could be extended in a number of directions by altering the linearly realised symmetries. For example, we could consider spontaneous breaking of Lorentz boosts as relevant for condensed matter physics and cosmology. Here the linearly realised symmetries would correspond to space-time translations and rotations. Systems of this type have been considered in \cite{SymmetricSuperfluids,Zoology}. We could also allow for extended SUSY. In that case at least one exceptional EFT is known which combines the full DBI (with a scalar and a gauge vector) with a VA fermion.  A simple generalisation of our analysis would allow one to confirm if this is the only possibility.

\subsection*{Acknowledgements}

It is a pleasure to thank Brando Bellazzini, Eric Bergshoeff, James Bonifacio, Scott Melville, Silvia Nagy and Antonio Padilla for very useful discussions. We thank Yusuke Yamada for participation in the early stages of this project. We acknowledge the Dutch funding agency ``Netherlands Organisation for Scientific Research'' (NWO) for financial support. 

\appendix
	
\section{Coset construction for supersymmetric Galileons} \label{Appendix}
As a concrete example of the general arguments we presented in section \ref{SuperspaceIH}, we now present the coset construction for supersymmetric Galileons. The bosonic sub-algebra is non-linearly realised by bi-Galileons with the coset construction worked out in \cite{GoonWess}. To this bosonic sub-algebra we add the appropriate fermionic generators for supersymmetrisation. All in all this algebra lives at level $n=1$ in the chiral superfield's superspace inverse Higgs tree and so the non-linear generators are $G$, $G_{\alpha \dot{\alpha}}$ and $S_{\alpha}$. The generators $G$ and $S_{\alpha}$ generate constant shift symmetries on the complex scalar and spin-$1/2$ fermion component fields respectively, while $G_{\alpha \dot{\alpha}}$ generates the Galileon symmetry on the complex scalar which is linear in the space-time coordinates. This is not an exceptional algebra but highlights the important parts of the SUSY coset construction. We remind the reader that an interesting Wess-Zumino term for this algebra appears in \cite{PhotonFarakos} while the soft amplitudes were discussed in \cite{SoftBootstrapSUSY,SuperGals}. For an interpretation of the algebra as an In\"onu-Wigner contraction, see \cite{InternalSUSY}. 
	
The only non-trivial commutators of the algebra are those required by the superspace inverse Higgs constraints. We have, in addition to the linearly realised super-Poincar\'{e} algebra and the commutators which define the Lorentz representation of the non-linear generators, 
\begin{align}
&\{Q_\mu, S_\nu\} = 2 \epsilon_{\mu \nu} G, \quad [P_{\mu \dot{\mu}},G_{\nu \dot{\nu}}] = i\epsilon_{\mu \nu} \epsilon_{\dot{\mu} \dot{\nu}}G, \quad [\bar{Q}_{\dot{\mu}},G_{\nu \dot{\nu}}] = i \epsilon_{\dot{\mu}\dot{\nu}} S_{\nu}\,.
 \end{align}
Introducing a Goldstone superfield for each non-linear generator, and including super-translations as usual, we parametrise the coset element for this symmetry breaking pattern as
\begin{align}
\Omega &= e^U e^V,
\end{align}
where 
\begin{align}
U &= \tfrac{i}{2}x^{\mu \dot{\mu}} P_{\mu \dot{\mu}} + i \theta^\mu Q_\mu + i \bar{\theta}_{\dot{\mu}} \bar{Q}^{\dot{\mu}}, \nonumber \\
V &= i \Phi G + i \bar{\Phi} \bar{G} + i \Psi^\mu S_\mu + i \bar{\Psi}_{\dot{\mu}}\bar{S}^{\dot{\mu}} -\tfrac{i}{2} \Lambda^{\mu \dot{\mu}} G_{\mu \dot{\mu}} - \tfrac{i}{2} \bar{\Lambda}^{\mu \dot{\mu}} \bar{G}_{\mu \dot{\mu}} \,.
\end{align}
Here the Greek letters from the middle of the alphabet ($\mu$, $\nu$, etc.) indicate space-time spinor indices (as opposed to the tangent space indices to be introduced in a moment).
	
The Maurer-Cartan form from which we can derive the superspace inverse Higgs constraints and the building blocks of invariant Lagrangians is given by
\begin{equation}
\omega = -i\Omega^{-1}d\Omega =-i e^{-V}(e^{-U}de^{U})e^V -i e^{-V}de^{V}.
\end{equation}
We begin by computing $e^{-U}de^U$ which, by using the SUSY algebra $\{Q_{\alpha},\bar{Q}_{\dot{\alpha}}\} = 2 P_{\alpha \dot{\alpha}}$, is given by
\begin{equation}
e^{-U}de^U = \tfrac{i}{2} P_{\mu \dot{\mu}} dx^{\mu \dot{\mu}} + i d\theta^\mu Q_\mu + i d\bar{\theta}_{\dot{\mu}} \bar{Q}^{\dot{\mu}} - P_{\mu \dot{\mu}}(d\theta^\mu \bar{\theta}^{\dot{\mu}} + d\bar{\theta}^{\dot{\mu}}\theta^\mu )\,.
\end{equation}
In the supersymmetric flat space basis, the exterior derivative is expressed as (see \cite{WessBagger} for more details)
\begin{equation}
d =  - \tfrac{1}{2} e^{\alpha \dot{\alpha}}\partial_{\alpha \dot{\alpha}} + e^\alpha D_\alpha + e_{\dot{\alpha}} \bar{D}^{\dot{\alpha}} \, , 
\end{equation}
so that each basis one-form $e^A$ multiplies a covariant object such that when $d$ acts on a superfield we get back another superfield. Note that $de^A \neq 0$ in general. Expressing $e^{-U}de^U$ in terms of these basis one-forms, we obtain
\begin{equation}
e^{-U}de^U = \tfrac{i}{2} e^{\alpha \dot{\alpha}}  P_{\alpha \dot{\alpha}} + ie^\alpha Q_\alpha + ie_{\dot{\alpha}}\bar{Q}^{\dot{\alpha}} \,.
\end{equation}
It is then simple to show that
\begin{align}
e^{-V}(e^{-U}de^{U})e^{V} &= \tfrac{i}{2}e^{\alpha\dot{\alpha}} P_{\alpha \dot{\alpha}}+ ie^{\alpha}Q_{\alpha} + ie_{\dot{\alpha}}\bar{Q}^{\dot{\alpha}} + (2e^{\alpha}\Psi_{\alpha} + \tfrac{i}{4}e_{\alpha \dot{\alpha}}\Lambda^{\alpha \dot{\alpha}})G \nonumber \\  &+  (2e_{\dot{\alpha}}\bar{\Psi}^{\dot{\alpha}} + \tfrac{i}{4}e_{\alpha \dot{\alpha}}\bar{\Lambda}^{\alpha \dot{\alpha}})\bar{G} + \tfrac{i}{2}e_{\dot{\beta}}\Lambda^{\beta \dot{\beta}}S_{\beta} -\tfrac{i}{2} e^{\alpha}\bar{\Lambda}_{\alpha \dot{\alpha}}\bar{S}^{\dot{\alpha}}.
\end{align}
The other part of the Maurer-Cartan form we need to compute is trivial since all non-linear generators commute amongst themselves. Indeed we have $e^{-V} d e^V = dV$. The full Maurer-Cartan form is then given by
\begin{align}
i\omega &= \tfrac{i}{2}e^{\alpha \dot{\alpha}}P_{\alpha \dot{\alpha}} + i e^\alpha Q_\alpha + ie_{\dot{\alpha}} \bar{Q}^{\dot{\alpha}} \nonumber  \\
&+ \bigg[ - \tfrac{i}{2}e_{\alpha \dot{\alpha}}(-\tfrac{1}{2}\Lambda^{\alpha \dot{\alpha}} +  \partial^{\alpha \dot{\alpha}}\Phi) + e^{\alpha}(2\Psi_{\alpha} + i D_{\alpha} \Phi)+ i e_{\dot{\alpha}}\bar{D}^{\dot{\alpha}}\Phi\bigg]G \nonumber  \\
& + \bigg[ - \tfrac{i}{2}e_{\alpha \dot{\alpha}}(-\tfrac{1}{2}\bar{\Lambda}^{\alpha \dot{\alpha}} +  \partial^{\alpha \dot{\alpha}}\bar{\Phi}) + ie^{\alpha} D_{\alpha} \bar{\Phi}+e_{\dot{\alpha}}(2 \bar{\Psi}^{\dot{\alpha}} + i\bar{D}^{\dot{\alpha}}\bar{\Phi})\bigg]\bar{G} \nonumber  \\
& + \bigg[-\tfrac{i}{2} e_{\alpha \dot{\alpha}} \partial^{\alpha \dot{\alpha}} \Psi^{\beta} +i e^\alpha D_\alpha \Psi^\beta  + e_{\dot{\beta}} \big(\tfrac{i}{2}\Lambda^{\beta \dot{\beta}} + i\bar{D}^{\dot{\beta}}\Psi^{\beta}\big)\bigg] S_\beta \nonumber  \\
& + \bigg[-\tfrac{i}{2} e_{\alpha \dot{\alpha}} \partial^{\alpha \dot{\alpha}} \bar{\Psi}^{\dot{\beta}} + e^{\beta}(\tfrac{i}{2}\bar{\Lambda}_{\beta}{}^{\dot{\beta}} + i D_\beta\bar{\Psi}^{\dot{\beta}})  + i e_{\dot{\alpha}}D^{\dot{\alpha}}\bar{\Psi}^{\dot{\beta}}\bigg] \bar{S}_{\dot{\beta}} \nonumber  \\
& +\bigg[\tfrac{i}{4}e^{\alpha \dot{\alpha}} \partial_{\alpha \dot{\alpha}} \Lambda^{\beta\dot{\beta}} -\tfrac{i}{2}e^\alpha D_\alpha \Lambda^{\beta \dot{\beta}}- \tfrac{i}{2}e_{\dot{\alpha}}\bar{D}^{\dot{\alpha}}\Lambda^{\beta \dot{\beta}}\bigg] G_{\beta \dot{\beta}} \nonumber  \\
& +\bigg[\tfrac{i}{4}e^{\alpha \dot{\alpha}} \partial_{\alpha \dot{\alpha}} \bar{\Lambda}^{\beta\dot{\beta}} -\tfrac{i}{2}e^\alpha D_\alpha \bar{\Lambda}^{\beta \dot{\beta}}- \tfrac{i}{2}e_{\dot{\alpha}}\bar{D}^{\dot{\alpha}}\bar{\Lambda}^{\beta \dot{\beta}}\bigg] \bar{G}_{\beta \dot{\beta}}\,.
\end{align}
Now as we mentioned in the main body, the coset covariant derivatives come from the product of the supervielbein and the Maurer-Cartan components, and since here the supervielbein is trivial we can simply read off the full coset covariant derivatives $\hat{D}_A$. The ones relevant for the superspace inverse Higgs constraints are
 \begin{align}
\hat{D}_{\mu \dot{\mu}} \Phi = \partial_{\mu \dot{\mu}} \Phi - \tfrac{1}{2} \Lambda_{\mu \dot{\mu}}, \quad \hat{D}_\mu \Phi = D_\mu \Phi - 2 i \Psi_\mu, \quad \bar{\hat{D}}^{\dot{\mu}} \Psi^\nu = \bar{D}^{\dot{\mu}} \Psi^\nu + \tfrac{1}{2}\Lambda^{\dot{\mu} \nu},
 \end{align}
which when set to zero yield the solutions
\begin{align}
\hat{D}_{\mu \dot{\mu}} \Phi = 0 \rightarrow \Lambda_{\mu \dot{\mu}} = 2 \partial_{\mu \dot{\mu}} \Phi, \quad \hat{D}_\mu \Phi = 0 \rightarrow 2\Psi_\mu = -i D_\mu \Phi, \quad \bar{\hat{D}}^{\dot{\mu}} \Psi^\nu = 0 \rightarrow \Lambda^{\nu \dot{\mu}} =-2\bar{D}^{\dot{\mu}} \Psi^\nu \, .
   \end{align}
Upon inserting these solutions back into the Maurer-Cartan form, we have the building blocks of invariant Lagrangians. 

Turning to the chirality condition for the superfield, from these solutions we find $\bar{D}_{\dot{\mu}}D_{\mu}\Phi = -2i \partial_{\mu \dot{\mu}} \Phi$ which, given the algebra of ordinary $\mathcal{N} = 1$ covariant derivatives $\{ D_\mu , \bar{D}_{\dot{\mu}} \} = -2 i \partial_{\mu \dot \mu}$, we find that we must have $\bar{D}_{\dot{\mu}}\Phi = 0$. This is a covariant condition since the ordinary barred spinor covariant derivative $\bar{D}_{\dot{\mu}}\Phi$ coincides with the hatted version $\hat{\bar{D}}_{\dot{\mu}} \Phi$. This is precisely the irreducibility condition for the chiral superfield which we see as a consistency condition following from imposing the superspace inverse Higgs constraints.


\begin{thebibliography}{99}

\bibitem{ScatteringAmplitudes}
  H.~Elvang and Y.~t.~Huang,
  arXiv:1308.1697 [hep-th].

\bibitem{SoftLimits1}
  C.~Cheung, K.~Kampf, J.~Novotny and J.~Trnka,
  Phys.\ Rev.\ Lett.\  {\bf 114} (2015) no.22,  221602
  [arXiv:1412.4095 [hep-th]].
  
\bibitem{SoftLimits2}
  C.~Cheung, K.~Kampf, J.~Novotny, C.~H.~Shen and J.~Trnka,
  JHEP {\bf 1702} (2017) 020
  [arXiv:1611.03137 [hep-th]].

\bibitem{SoftLimits3}
  A.~Padilla, D.~Stefanyszyn and T.~Wilson,
  JHEP {\bf 1704} (2017) 015
  [arXiv:1612.04283 [hep-th]].
  
\bibitem{Recursion}
  C.~Cheung, K.~Kampf, J.~Novotny, C.~H.~Shen and J.~Trnka,
  Phys.\ Rev.\ Lett.\  {\bf 116} (2016) no.4,  041601
  [arXiv:1509.03309 [hep-th]].
  
\bibitem{LowYin}
  I.~Low and Z.~Yin,
  arXiv:1904.12859 [hep-th].
  
\bibitem{SoftLimitVector}
  C.~Cheung, K.~Kampf, J.~Novotny, C.~H.~Shen, J.~Trnka and C.~Wen,
  Phys.\ Rev.\ Lett.\  {\bf 120} (2018) no.26,  261602
  [arXiv:1801.01496 [hep-th]].
  
\bibitem{SoftBootstrapSUSY}
  H.~Elvang, M.~Hadjiantonis, C.~R.~T.~Jones and S.~Paranjape,
  arXiv:1806.06079 [hep-th].
  
    
\bibitem{SuperGals}
  H.~Elvang, M.~Hadjiantonis, C.~R.~T.~Jones and S.~Paranjape,
  Phys.\ Lett.\ B {\bf 781} (2018) 656
  [arXiv:1712.09937 [hep-th]].
  
\bibitem{Recursion1}
  R.~Britto, F.~Cachazo and B.~Feng,
  Nucl.\ Phys.\ B {\bf 715} (2005) 499
  [hep-th/0412308].
  
\bibitem{Recursion2}
  R.~Britto, F.~Cachazo, B.~Feng and E.~Witten,
  Phys.\ Rev.\ Lett.\  {\bf 94} (2005) 181602
  [hep-th/0501052].
  
\bibitem{SimplestQFT}
  N.~Arkani-Hamed, F.~Cachazo and J.~Kaplan,
  JHEP {\bf 1009} (2010) 016
  [arXiv:0808.1446 [hep-th]].


\bibitem{Adler1}
  S.~L.~Adler,
  Phys.\ Rev.\  {\bf 137} (1965) B1022.
  
\bibitem{Adler2}
  S.~L.~Adler,
  Phys.\ Rev.\  {\bf 139} (1965) B1638.
  
\bibitem{Mapping}
  P.~Creminelli, M.~Serone and E.~Trincherini,
  JHEP {\bf 1310} (2013) 040
  [arXiv:1306.2946 [hep-th]].
  
\bibitem{dilaton1}
  R.~H.~Boels and W.~Wormsbecher,
  arXiv:1507.08162 [hep-th].
  
\bibitem{dilaton2}
  P.~Di Vecchia, R.~Marotta, M.~Mojaza and J.~Nohle,
  Phys.\ Rev.\ D {\bf 93} (2016) no.8,  085015
  [arXiv:1512.03316 [hep-th]].
  
\bibitem{dilaton3}
  M.~Bianchi, A.~L.~Guerrieri, Y.~t.~Huang, C.~J.~Lee and C.~Wen,
  JHEP {\bf 1610} (2016) 036
  [arXiv:1605.08697 [hep-th]].
  
\bibitem{Internal1}
  S.~R.~Coleman, J.~Wess and B.~Zumino,
  Phys.\ Rev.\  {\bf 177} (1969) 2239.

\bibitem{Internal2}
  C.~G.~Callan, Jr., S.~R.~Coleman, J.~Wess and B.~Zumino,
  Phys.\ Rev.\  {\bf 177} (1969) 2247.

\bibitem{Spacetime1}
  D.~V.~Volkov,
  Fiz.\ Elem.\ Chast.\ Atom.\ Yadra {\bf 4} (1973) 3.

\bibitem{Spacetime2}
  E.~A.~Ivanov and V.~I.~Ogievetsky,
  Teor.\ Mat.\ Fiz.\  {\bf 25} (1975) 164.
  
\bibitem{LieAlgebraicScalar1}
  M.~P.~Bogers and T.~Brauner,
  JHEP {\bf 1805} (2018) 076
  [arXiv:1803.05359 [hep-th]].
  
\bibitem{LieAlgebraicScalar2}
  M.~P.~Bogers and T.~Brauner,
  Phys.\ Rev.\ Lett.\  {\bf 121} (2018) no.17,  171602
  [arXiv:1802.08107 [hep-th]].
  
\bibitem{LieAlgebraicVector}
  R.~Klein, E.~Malek, D.~Roest and D.~Stefanyszyn,
  Phys.\ Rev.\ D {\bf 98} (2018) no.6,  065001
  [arXiv:1806.06862 [hep-th]].
  
\bibitem{RSW}
  D.~Roest, D.~Stefanyszyn and P.~Werkman,
  arXiv:1903.08222 [hep-th].
  
  
\bibitem{AdS}
  J.~Bonifacio, K.~Hinterbichler, A.~Joyce and R.~A.~Rosen,
  arXiv:1812.08167 [hep-th].
  
\bibitem{SymmetricSuperfluids}
  E.~Pajer and D.~Stefanyszyn,
  arXiv:1812.05133 [hep-th].

  
  
\bibitem{Goldstone}
  J.~Goldstone, A.~Salam and S.~Weinberg,
  Phys.\ Rev.\  {\bf 127} (1962) 965.
  
  
\bibitem{LowManohar}
  I.~Low and A.~V.~Manohar,
  Phys.\ Rev.\ Lett.\  {\bf 88} (2002) 101602
  [hep-th/0110285].
  
  
\bibitem{ExtendedShift}
K.~Hinterbichler and A.~Joyce,
Int.\ J.\ Mod.\ Phys.\ D {\bf 23} (2014) no.13,  1443001
[arXiv:1404.4047 [hep-th]].


\bibitem{Galileon}
  A.~Nicolis, R.~Rattazzi and E.~Trincherini,
  Phys.\ Rev.\ D {\bf 79} (2009) 064036
  [arXiv:0811.2197 [hep-th]].

  
\bibitem{WZ}
  G.~Goon, K.~Hinterbichler, A.~Joyce and M.~Trodden,
  JHEP {\bf 1206} (2012) 004
  [arXiv:1203.3191 [hep-th]].


  
\bibitem{BI}
  M.~Born and L.~Infeld,
  Proc.\ Roy.\ Soc.\ Lond.\ A {\bf 144} (1934) no.852,  425.
  
\bibitem{Dirac}
  P.~A.~M.~Dirac,
  Proc.\ Roy.\ Soc.\ Lond.\ A {\bf 268} (1962) 57.
  

  
\bibitem{SpecialGalileon}
  K.~Hinterbichler and A.~Joyce,
  Phys.\ Rev.\ D {\bf 92} (2015) no.2,  023503
  [arXiv:1501.07600 [hep-th]].
  
 
\bibitem{DoubleCopy}
  M.~Carrillo González, R.~Penco and M.~Trodden,
  JHEP {\bf 1811} (2018) 065
  [arXiv:1809.04611 [hep-th]].
  
\bibitem{GeometryGal}
  J.~Novotny,
  Phys.\ Rev.\ D {\bf 95} (2017) no.6,  065019
  [arXiv:1612.01738 [hep-th]].
  
\bibitem{VA}
  D.~V.~Volkov and V.~P.~Akulov,
  Phys.\ Lett.\  {\bf 46B} (1973) 109.
  
\bibitem{FermionSoft1}
  R.~Kallosh, A.~Karlsson and D.~Murli,
  JHEP {\bf 1703} (2017) 081
  [arXiv:1609.09127 [hep-th]].

\bibitem{FermionSoft2}
  B.~Bellazzini,
  JHEP {\bf 1702} (2017) 034
  [arXiv:1605.06111 [hep-th]].
  
  \bibitem{IvanovKapustnikov}
  E.~A.~Ivanov and A.~A.~Kapustnikov,
  J.\ Phys.\ A {\bf 11} (1978) 2375.
  
  \bibitem{IvanovKapustnikov2}
  E.~A.~Ivanov and A.~A.~Kapustnikov,
  J.\ Phys.\ G {\bf 8} (1982) 167.
  
\bibitem{UV-InverseHiggs}
  S.~Endlich, A.~Nicolis and R.~Penco,
  Phys.\ Rev.\ D {\bf 89} (2014) no.6,  065006
  [arXiv:1311.6491 [hep-th]].
  
\bibitem{Ward1}
  M.~T.~Grisaru, H.~N.~Pendleton and P.~van Nieuwenhuizen,
  Phys.\ Rev.\ D {\bf 15} (1977) 996.
  
\bibitem{Ward2}
  M.~T.~Grisaru and H.~N.~Pendleton,
  Nucl.\ Phys.\ B {\bf 124} (1977) 81.
  
\bibitem{Ward3}
  H.~Elvang, D.~Z.~Freedman and M.~Kiermaier,
  JHEP {\bf 1010} (2010) 103
  [arXiv:0911.3169 [hep-th]].
  
    
\bibitem{WessBagger}  
J. Wess and J. Bagger ``Supersymmetry and Supergravity", \textit{Princeton Series in Physics} (1992)
  
    \bibitem{Olver}
P.J. Olver, ``Applications of Lie Groups to Differential Equations", \textit{Springer} (1986)

\bibitem{KRS}
  R.~Klein, D.~Roest and D.~Stefanyszyn,
  JHEP {\bf 1710} (2017) 051
  [arXiv:1709.03525 [hep-th]].
  
\bibitem{McArthur}
  I.~N.~McArthur,
  JHEP {\bf 1011} (2010) 140
  [arXiv:1009.3696 [hep-th]].


\bibitem{GoonWess}
  G.~Goon, K.~Hinterbichler, A.~Joyce and M.~Trodden,
  JHEP {\bf 1206} (2012) 004
  [arXiv:1203.3191 [hep-th]].
  
\bibitem{Wheel}
  L.~V.~Delacrétaz, S.~Endlich, A.~Monin, R.~Penco and F.~Riva,
  JHEP {\bf 1411} (2014) 008
  [arXiv:1405.7384 [hep-th]].
  
\bibitem{BaggerGalperinScalar}
  J.~Bagger and A.~Galperin,
  Phys.\ Lett.\ B {\bf 336} (1994) 25
  [hep-th/9406217].
  
\bibitem{BaggerGalperinMaxwell} 
  J.~Bagger and A.~Galperin,
  Phys.\ Rev.\ D {\bf 55}, 1091 (1997)
  [hep-th/9608177].
    
  
\bibitem{IvanovBellucciKrivonos}
  S.~Bellucci, E.~Ivanov and S.~Krivonos,
  Nucl.\ Phys.\ B {\bf 672} (2003) 123
  [hep-th/0212295].
  
  
  
\bibitem{LehnersKhoury1}
  J.~Khoury, J.~L.~Lehners and B.~Ovrut,
  Phys.\ Rev.\ D {\bf 83} (2011) 125031
  [arXiv:1012.3748 [hep-th]].


\bibitem{InternalSUSY}
  D.~Roest, P.~Werkman and Y.~Yamada,
  JHEP {\bf 1805} (2018) 190
  [arXiv:1710.02480 [hep-th]].

\bibitem{multiDBI}
  K.~Hinterbichler, M.~Trodden and D.~Wesley,
  Phys.\ Rev.\ D {\bf 82} (2010) 124018
  [arXiv:1008.1305 [hep-th]].
  
\bibitem{bi-gal}
  A.~Padilla, P.~M.~Saffin and S.~Y.~Zhou,
  JHEP {\bf 1012} (2010) 031
  [arXiv:1007.5424 [hep-th]].
  

      \bibitem{RocekTseytlin}
  M.~Rocek and A.~A.~Tseytlin,
  Phys.\ Rev.\ D {\bf 59} (1999) 106001
  [hep-th/9811232].
  
\bibitem{PhotonFarakos}
  F.~Farakos, C.~Germani and A.~Kehagias,
  JHEP {\bf 1311} (2013) 045
  [arXiv:1306.2961 [hep-th]].
  
  
\bibitem{VectorNoGo}
  C.~Deffayet, A.~E.~Gümrükçüoğlu, S.~Mukohyama and Y.~Wang,
  JHEP {\bf 1404} (2014) 082
  [arXiv:1312.6690 [hep-th]].
  
  \bibitem{BaggerGalperinLinear}
  J.~Bagger and A.~Galperin,
  Phys.\ Lett.\ B {\bf 412} (1997) 296
  doi:10.1016/S0370-2693(97)01030-7
  [hep-th/9707061].


  

  


\bibitem{Equivalence}
  S.~Bellucci, E.~Ivanov and S.~Krivonos,
  Phys.\ Rev.\ D {\bf 66} (2002) 086001
   Erratum: [Phys.\ Rev.\ D {\bf 67} (2003) 049901]
  [hep-th/0206126].

  
\bibitem{IvanovBellucciKrivonos2}
  S.~Bellucci, E.~Ivanov and S.~Krivonos,
  Phys.\ Lett.\ B {\bf 558} (2003) 182
  [hep-th/0212142].
    

\bibitem{DeenOvrut}
  R.~Deen and B.~Ovrut,
  JHEP {\bf 1708} (2017) 014
  [arXiv:1705.06729 [hep-th]].
  
\bibitem{AchucarroTownsend}
  A.~Achucarro, J.~M.~Evans, P.~K.~Townsend and D.~L.~Wiltshire,
  Phys.\ Lett.\ B {\bf 198} (1987) 441.
  
\bibitem{SuperspaceGeometry}
  J.~A.~De Azcarraga and P.~K.~Townsend,
  Phys.\ Rev.\ Lett.\  {\bf 62} (1989) 2579.
  
\bibitem{Zoology}
  A.~Nicolis, R.~Penco, F.~Piazza and R.~Rattazzi,
  JHEP {\bf 1506} (2015) 155
  doi:10.1007/JHEP06(2015)155
  [arXiv:1501.03845 [hep-th]].
  

  
 \end{thebibliography}
\end{document}